\documentclass[a4paper,11pt]{article}
\usepackage{amsmath,amssymb,bm,epsfig,color,cite,braket,mathrsfs,slashed,multirow,graphicx,latexsym,cancel,comment}

\renewcommand{\thefootnote}{\fnsymbol{footnote}}

\newcommand{\D}{{\rm d}}

\usepackage{hyperref,xcolor}
\hypersetup{
colorlinks=true,
linktocpage=true,
linkcolor=blue,
filecolor=blue,      
urlcolor=blue,
citecolor=blue,
   }

\numberwithin{equation}{section}

\usepackage{siunitx}

\usepackage{tikz}

\begin{document}

\title{
{\Large \bf Peaks sphericity of non-Gaussian random fields\\*[15pt]}} 

\author{Cristiano Germani$^{1}$,\footnote{\href{mailto:germani@icc.ub.edu}{germani@icc.ub.edu}}\ \
Mohammad Ali Gorji$^{2}$,\footnote{
\href{mailto:gorji@ibs.re.kr}{gorji@ibs.re.kr}}\ \ 
Michiru Uwabo-Niibo$^{2,3}$,\footnote{
\href{mailto:g2370609@edu.cc.ocha.ac.jp}{g2370609@edu.cc.ocha.ac.jp}}\ \ 
Masahide Yamaguchi$^{2,4,5}$\footnote{
\href{mailto:gucci@ibs.re.kr}{gucci@ibs.re.kr}}
\\*[15pt]
$^1${\it \small
Departament de F\'{i}sica Qu\`{a}ntica i Astrof\'{i}sica and Institut de Ci\`{e}ncies del Cosmos, Universitat de Barcelona,}
\\
{\it \small Mart\'{i} i Franqu\`{e}s 1, 08028 Barcelona, Spain}
\\*[5pt]
$^2${\it\small
Cosmology, Gravity, and Astroparticle Physics Group, Center for
Theoretical Physics of the Universe, }\\{\it\small  Institute for Basic Science (IBS), Daejeon,
34126, Korea} 
\\*[5pt]
$^3${\it \small
Department of Physics, Graduate School of Humanities and Sciences,}\\{\it\small  Ochanomizu University, Tokyo 112-8610, Japan} 
\\
$^{4}${\it \small
Department of Physics, Institute of Science Tokyo,}\\{\it \small
2-12-1 Ookayama, Meguro-ku, Tokyo 152-8551, Japan}
\\
$^{5}${\it \small
Department of Physics and IPAP,
Yonsei University, 50 Yonsei-ro, Seodaemun-gu, Seoul 03722, Korea}
\\*[50pt]
}

\date{
\centerline{\small \bf Abstract}
\begin{minipage}{0.9\linewidth}
\medskip \medskip \small 
We formulate the statistics of peaks of non-Gaussian random fields and implement it to study the sphericity of peaks. For non-Gaussianity of the local type, we present a general formalism valid regardless of how large the deviation from Gaussian statistics is. For general types of non-Gaussianity, we provide a framework that applies to any system with a given power spectrum and the corresponding bispectrum in the regime in which contributions from higher-order correlators can be neglected. We present an explicit expression for the most probable values of the sphericity parameters, including the effect of non-Gaussianity on the shape. We show that the effects of small perturbative non-Gaussianity on the sphericity parameters are negligible, as they are even smaller than the subleading Gaussian corrections. In contrast, we find that large non-Gaussianity can significantly distort the peak configurations, making them much less spherical.
\end{minipage}
}

\maketitle{}
\thispagestyle{empty}
\addtocounter{page}{-1}
\clearpage
\noindent
\hrule
\tableofcontents
\noindent
\hrulefill

\renewcommand{\thefootnote}{\arabic{footnote}}
\setcounter{footnote}{0}

\newpage

\section{Introduction}

Peak theory, established in Ref.~\cite{Bardeen:1985tr} for Gaussian variables, is a comprehensive mathematical analysis of peaks of random fields, with wide applications in physics. One prominent example is cosmological structure formation. 
Due to gravitational instability, the primordial density perturbations evolve to form the observed large-scale structures in the Universe. Though many candidates of the source of these primordial perturbations have been proposed, one of the powerful candidates is inflation, during which the seeds of the observed large-scale structures in the Universe are generated \cite{Mukhanov:2005sc,Weinberg:2008zzc}. The small inhomogeneous perturbations, on top of the homogeneous and isotropic Universe, are generated from inflationary quantum fluctuations. These perturbations are stretched to superhorizon scales during inflation, re-enter the horizon after inflation, and finally evolve to form the observed large-scale structures in the Universe. This is the standard scenario for generating the primordial inhomogeneities. Regardless of the source, the key quantity that characterizes the perturbations might be expressed in terms of the random energy density contrast, $\delta(\vec{r},t) = (\rho(\vec{r},t)-\langle \rho(t)\rangle )/\langle \rho(t)\rangle$, where $\langle \rho(t)\rangle $ is the ensemble average of the energy density over the whole Universe in the spatially flat gauge. 

Peak theory provides the statistical prediction of the properties of peaks of a random field, such as their number density, correlations in position space, shapes, peculiar velocities, and so on. Because at linear regime $\delta(\vec{r},t)$ is Gaussian, peak theory has then played an important role in predicting the initial conditions of structure formation since its establishment (for a review, see e.g.~\cite{Cooray:2002dia}).

On the other hands, non-Gaussianities may play an essential role on rare processes. For example, primordial black holes (PBH) may be the result of gravitational collapses of rarely large curvature perturbations seeded by inflation. Thus, even a small non-Gaussianity, may change the prediction of the PBHs abundance (see e.g.\cite{ Young:2013oia,Franciolini:2018vbk,Atal:2018neu, Yoo:2019pma, Kitajima:2021fpq, Young:2022phe}). Moreover, because the local gravitational potential - the key object responsible for the collapse \cite{Shibata:1999zs} - is non-linearly related to curvature perturbations, the statistics of PBH would be non-Gaussian \cite{Germani:2019zez,Germani:2023ojx} even if the perturbations were Gaussian distributed. Ignoring this fact generically leads to large errors in the prediction of PBHs abundance \cite{Fumagalli:2024kxe}. 

There is however an extra aspect that has been so far overlooked. Numerical \cite{Musco:2018rwt,Escriva:2019nsa} and analytical \cite{Escriva:2019phb,Escriva:2020tak} methods to find the threshold for the gravitational potential (the so-called compaction function), assume a spherically symmetric gravitational collapse (or a weakly non-spherical shapes \cite{Escriva:2024lmm,Escriva:2024aeo}). This is a fairly good assumption for a Gaussianly distributed rarely large random variable \cite{Bardeen:1985tr}, but it might not be such for a non-Gaussian one, as it is the compaction function. 

In this paper we shall investigate this point in a general manner. In other words, we shall ask the question of whether non-Gaussianities can largely change the shape of rare perturbations with respect to their Gaussian counterpart:
if non-Gaussianity affects the property of the peaks, the assumption of spherical symmetry may indeed be a bad approximation. For example, consider a local-type non-Gaussian random field in the position space $\vec{r}$ defined as $F(\vec{r}) = F_{G}(\vec{r}) + f_{\rm NL}F_{G}(\vec{r})^{2}, \label{eq:local-type non-G with shift}$ where $F_{G}$ obeys Gaussian statistics. 
There, high peaks are more probable than those of the Gaussian case for positive $f_{\rm NL}$. Thus, 
the height of peaks may have nothing to do with the rareness of peaks and hence the assumption of spherical symmetry for high peaks may be violated. Furthermore, even this kind of argument may not necessarily hold true for general types of non-Gaussianity.

While other authors have already extended peak theory to the non-Gaussian case (e.g.~Refs.~\cite{Gay:2011wz,Young:2013oia,Lazeyras:2015giz,Franciolini:2018vbk,Atal:2018neu,Yoo:2019pma,Matsubara:2020lyv,Kitajima:2021fpq,Riccardi:2021rlf,Young:2022phe}) those papers mainly focused on the number of peaks without addressing the shape of the configurations, such as deviations from spherical symmetry. Although non-Gaussianity and non-sphericity are in principle different notions, careful consideration is necessary when they are taken into account separately. To the best of our knowledge, no study has yet provided a general framework that accounts for both non-Gaussianity and non-sphericity effects. In this paper, for the first time, we provide the effects of non-Gaussianity on the non-sphericity from the statistical point of view.

The paper is organized as follows. In Sec.~\ref{sec:Gaussian}, we review peak theory of a Gaussian random field according to Ref.~\cite{Bardeen:1985tr}. In this section, the definitions of sphericity parameters, ellipticity and prolateness, are given.
In Sec.~\ref{sec:exact local nonG}, we show the probability distribution of sphericity parameters in the case of local-type non-Gaussianity, without specifying the functional form of $F(\vec{r})=F[F_{G}(\vec{r})]$.  
We discuss sphericity of peaks with general bispectrum in Sec.~\ref{sec:non-Gaussian peaks}. In Sec.~\ref{subsec:general edgeworth}, we review Edgeworth expansion which is a mathematical technique to construct a probability distribution function (PDF) of a random variable out of its cumulants. In Sec.~\ref{sec:general nonG}, assuming small deviations from Gaussian statistics, we use Edgeworth expansion to construct a general setup applicable to any system with a given power spectrum and the corresponding bispectrum, while neglecting contributions from higher-order correlators. In Sec.~\ref{sec-tail}, we focus on the tail of the PDF and show that large non-Gaussianity can make higher peaks tend to be much less spherical. Sec.~\ref{sec:summary} is devoted to the summery and conclusions, while additional details of the derivation are given in appendices, \ref{app-xi-diag}, \ref{app-functions}, \ref{app-edgeworth} and \ref{app-I-ABC}. For validation and consistency checking, in Appendix~\ref{sec:specific bispectrums}, we confirmed that the different formalisms that we have developed in Sec.~\ref{sec:general nonG} and Sec.~\ref{sec:exact local nonG} yield the same result for a specific local-type non-Gaussianity.

\section{Peaks in Gaussian random fields}\label{sec:Gaussian}
In this section, we review the statistical sphericity of peaks of a Gaussian random field following Ref.~\cite{Bardeen:1985tr}. Readers who are familiar with this subject may wish to directly move to the next section.

For a random field $F(\vec{r})$, the variables that characterize the peak are the field itself and its spatial derivatives $\nabla F(\vec{r}),\nabla\nabla F(\vec{r}),\cdots$. To study the sphericity of the peak surroundings we consider up to the second derivatives of the field
\begin{align}
\eta_{i} \equiv \nabla_{i}F \,,
\qquad
\xi_{ij} \equiv \nabla_{i}\nabla_{j}F \,,
\end{align}
where clearly $\xi_{ij}$ is symmetric. Therefore we deal with the following set of statistical variables
\begin{align}\label{h-def}
h_A =\{F,\eta_i,\xi_{ij}\} \,, 
\qquad 
A = 1, \cdots, 10 \,.
\end{align}
We treat the above variables as independent. Considering Gaussian statistics, the corresponding 10-dimensional joint PDF takes the form
\begin{align}\label{PDF}
&P\left(F,\eta,\xi\right)\D F\D^3\eta \D^6\xi =\tilde{N}e^{-Q\left(F,\eta,\xi\right)}\D F\D^3\eta \D^6\xi \,,
\end{align}
where
\begin{align}\label{Q-def}
&Q\left(F,\eta,\xi\right)=\frac{1}{2}\sum_{AB}h_AM^{-1}_{AB} h_{B} \,,
&&M_{AB}\equiv\langle h_{A}h_{B}\rangle \,.
\end{align}
The numerical constant $\tilde{N}\propto ({\rm det}M)^{-1/2}$ should be fixed such that $\int{P}\left(F,\eta,\xi\right)\D F\D^3\eta \D^6\xi=1$. In order to find the components of $M_{AB}$, we define the Fourier transform of $F(\vec{r})$ as
\begin{align}
F(\vec{k}) = \int F(\vec{r})e^{-i\vec{k}\cdot \vec{r}}\D^{3}r \,.
\end{align}
In the homogeneous and isotropic Universe, the power spectrum in Fourier space is defined as
\begin{align}
    \langle F(\vec{k})F(\vec{k}')\rangle = (2\pi)^{3}\delta^{(3)}(\vec{k} + \vec{k}')P(k),\label{eq:definition of power-spectrum}
\end{align}
where $k=|\vec{k}|$. We assume $\langle F(\vec{k})\rangle = 0$ i.e.~$\langle F(\vec{r})\rangle = 0$. Note that if $\langle F(\vec{k})\rangle \neq  0$ i.e.~$\langle h_{A}(\vec{r})\rangle \neq 0$, the exponential factor is modified by $h_{A}(\vec{r}) \rightarrow \tilde{h}_{A}(\vec{r}) = h_{A}(\vec{r})-\langle h_{A}(\vec{r})\rangle$ and $M_{AB}(\vec{r})\rightarrow \tilde{M}_{AB} (\vec{r})= \langle \tilde{h}_{A}(\vec{r})\tilde{h}_{B}(\vec{r})\rangle$.

As the PDF \eqref{PDF} is (multi-)Gaussian, the ensemble average of any products of $F(\vec{r})$, $\eta$ and $\xi$ are characterized by the power spectrum $P(k)$. It is then very useful to define the correlation parameters 
\begin{align}\label{sigma-def}
    \sigma_{j}^{2} \equiv \int \frac{d^{3}k}{(2\pi)^{3}}P(k)k^{2j} \,.
\end{align}
With straightforward calculations, we find that the only nonzero components of $10\times10$ matrix $M_{AB}$ are
\begin{align}\begin{split}
&M_{FF} = \sigma_{0}^{2},\quad M_{F\xi_{ij}} = -M_{\eta_{i}\eta_{j}} = -\frac{1}{3}\sigma_{1}^{2}\delta_{ij},\\
&M_{\xi_{ij}\xi_{kl}} =\frac{1}{15}\sigma_{2}^{2}(\delta_{ij}\delta_{kl} + \delta_{ik}\delta_{jl} + \delta_{il}\delta_{jk})\,.
\label{eq:2 pt correlation}
\end{split}\end{align}
Substituting the above results in \eqref{Q-def}, we can easily find the explicit form of $Q$ in terms of $(F,\eta_i,\xi_{ij})$ that is given by \eqref{Q-Trace}. It turns out that only ${\rm Tr}(\xi)$ and ${\rm Tr}\left(\xi^2\right)$ show up in $Q$. Consequently, we can diagonalize $\xi_{ij}$ and express $Q$ in terms of $(F,\eta_i)$ and three eigenvalues of $\xi_{ij}$ as shown in Eq. \eqref{Q-Lambda}. As the other three independent components of $\xi_{ij}$ do not appear in $Q$, we can integrate them out in the measure (see Eq.~\eqref{measure-Lambda} and the paragraph after that in appendix \ref{app-xi-diag} for the details) and the PDF \eqref{PDF} takes the form
\begin{align}\label{PDF-lambda}
&P\left(\nu,\alpha,\varsigma\right) \D \nu\D^3\alpha \D^6\varsigma 
= \frac{\pi^2}{3} N e^{-Q\left(\nu,\alpha,\lambda\right)} |(\lambda_{1}-\lambda_{2})(\lambda_{2}-\lambda_{3})(\lambda_{3}-\lambda_{1})|
\D \nu\D^3\alpha \D\lambda_{1}\D\lambda_{2}\D\lambda_{3} \,,
\end{align}
where $N=\tilde{N}\sigma_{0}\sigma_{1}^{3}\sigma_{2}^{6}$ and
\begin{align}\label{Q-lambda}
2Q\left(\nu,\alpha,\lambda\right) =& \frac{1}{1-\gamma^{2}}
\Big( \nu^2 - 2 \gamma \nu\sum_i\lambda_i \Big)
+ \frac{5\gamma^2-3}{2(1-\gamma^2)} \big(\sum_i \lambda_i\big)^2 + \frac{15}{2}\sum_i\lambda_i^2
+ 3 \sum_{i}\alpha_{i}^{2} \,,
\end{align}
with 
\begin{align}\label{def-gamma}
\gamma\equiv \sigma_{1}^{2}/\sigma_{0}\sigma_{2} \,.
\end{align}
In \eqref{PDF-lambda}, we have defined normalized variables 
\begin{align}\label{h-Normalized}
\nu &\equiv \frac{F}{\sigma_0} \,,
\qquad 
\alpha_i \equiv \frac{\eta_i}{\sigma_1} \,,
\qquad 
\varsigma_{ij} \equiv \frac{\xi_{ij}}{\sigma_2} \,,
\end{align}
and $\lambda_i$ are eigenvalues of $\left(-\varsigma_{ij}\right)$. Note that  $\xi_{ij}=\nabla_i\nabla_jF$ is negative for a peak and that is why we have considered minus sign to deal with positive definite variables $\lambda_i$. 

Now, we suppose that $F(\vec{r})$ has a peak at $\vec{r}_{\rm pk}$. Expanding $F(\vec{r})$ around the peak, we find
\begin{align}
F(\vec{r})
=&F(\vec{r}_{\rm pk}) + \frac{1}{2}\sum_{i=1}^{3} \xi_{ij}  (r-r_{\rm pk})^{i} (r-r_{\rm pk})^{j} + \mathcal{O}(|\vec{r}-\vec{r}_{\rm pk}|^{3}) \,,
\label{eq:configuration around a peak}
\end{align}
where we have used the fact that $\eta_{i}(r_{\rm pk})=0$ since $r_{\rm pk}$ is a maximum. The above equation can be written as the equation of an ellipsoid  $\sum_{i}\left(R_{ij}r^{j}\right)^{2}/a_{i}^{2}=1$, where $R_{ij}$ are components of the matrix that diagonalizes $\xi_{ij}$ (defined in Eq. \eqref{trans}), with the semi-axes
\begin{align}
    a_{i}=\sqrt{\frac{2(F(\vec{r}_{\rm pk})-F(\vec{r}))}{\lambda_{i}\sigma_{2}}}.
\end{align}
We thus define the sphericity parameters as
\begin{align}
    \begin{split}
        e=\frac{\lambda_{1}-\lambda_{3}}{2\left|\sum_{i}\lambda_{i}\right|},
        \qquad p=\frac{\lambda_{1}-2\lambda_{2}+\lambda_{3}}{2\left|\sum_{i}\lambda_{i}\right|}.\label{eq:definition of e, p}
    \end{split}
\end{align}
These parameters are also referred as ellipticity and prolateness. The configuration becomes spherical when $\lambda_{1}=\lambda_{2}=\lambda_{3}$, i.e.~$e=p=0$. We see that the sphericity of a peak is determined by hierarchies of $\lambda_{i}$, i.e.~the eigenvalues of the second derivative of $F(\vec{r})$. This is consistent with the fact that to study the sphericity of a peak, we only need to keep up to the second derivatives.

Defining new variables
\begin{align}
x&\equiv \sum_{i}\lambda_{i} ,
\qquad 
y \equiv \frac{\lambda_{1}-\lambda_{3}}{2}=e|x|,
\qquad 
z \equiv \frac{\lambda_{1}-2\lambda_{2}+\lambda_{3}}{2}=p|x| \,,
\label{eq:definition of x,y,z}
\end{align}
the PDF turns out to be
\begin{align}
\begin{split}
P\left(\nu,\alpha,\varsigma\right) \D\nu \D^3\alpha \D^6\varsigma = &\frac{2}{3}\frac{2\pi^{2}}{3!} N e^{-Q\left(\nu,\alpha,x,y,z\right)}|2y(y^{2}-z^{2})| \D\nu \D^3\alpha \D{x} \D{y} \D{z} \,,
\label{eq:7 dim PDF of Gaussian (xyz)}
\end{split}
\end{align}
where
\begin{align}
2Q\left(\nu,\alpha,x,y,z\right)
= \frac{1}{1-\gamma^{2}} 
\left(
\nu^2 - 2\gamma \nu x + x^{2}
\right) + 5 \left( 3y^{2}+z^{2} \right) + 3 \sum_{i}\alpha_{i}^{2} \,.
\label{eq:Q0 of Gaussian}
\end{align}
Note that $\gamma$ characterizes the cross correlation of $\nu$ and $x$. If the power spectrum is a $\delta$-function, it is maximized to $\gamma=1$. If these two variables are fully independent, $\gamma=0$. 

Without loss of generality, we assume $\lambda_{1}\geq \lambda_{2}\geq \lambda_{3}$
which is equivalent to $y\geq |z|\geq 0$. We can then get rid of the absolute value symbol of $y(y^{2}-z^{2})$ and we find
\begin{align}\label{P-I}
\begin{split}
P\left(\nu,\alpha,\varsigma\right) \D\nu \D^3\alpha \D^6\varsigma = &\frac{8\pi^{2}}{3} N e^{-Q(\nu,\alpha,x,y,z)} y(y^{2}-z^{2})\chi(y,z) \D\nu \D^3\alpha \D{x} \D{y} \D{z} \,,
\end{split}
\end{align}
where
\begin{equation}\label{eq:definition of chi(y,z)}
    \chi(y,z) =
        \begin{cases}
        1 & (y\geq |z|\geq 0) \\
        0 & \text{otherwise} \,.
        \end{cases}
\end{equation}
The factor $3!$ in Eq.~\eqref{eq:7 dim PDF of Gaussian (xyz)} is canceled out by fixing the order of $\lambda_{i}$. Note that due to the restricted range of $y$ and $z$, they do not obey Gaussian statistics. 

One can consider either of positive (peaks) or negative (troughs) $x$. Here let us focus on $x>0$. We will discuss $x<0$ in Sec.~\ref{sec:valleys}.
The number of peaks $N_{\rm pk}$ can be expressed in terms of the number density $n_{\rm pk}\left( \vec{r} \right) = \sum_{{\rm pk}=1}^{N_{\rm pk}} \delta^{(3)}\left( \vec{r} - \vec{r}_{\rm pk} \right)$ such that $N_{\rm pk} = \int \D^3{r}\,n_{\rm pk} \left( \vec{r} \right)$. Using the fact that around the peak
\begin{align}
\eta_{i}(\vec{r}) = \sum_{j}\xi_{ij}(\vec{r}_{\rm pk})(r-r_{\rm pk})_{j} + \mathcal{O}(|\vec{r}-\vec{r}_{\rm pk}|^{2}) \,,
\end{align}
we find
\begin{align}
n_{\rm pk}(\vec{r})
&=  \theta(\lambda_{1})\theta(\lambda_{2})\theta(\lambda_{3}) |{\rm det}\xi| \delta^{(3)}(\vec{\eta}) \,,
\label{n-peak}
\end{align}
where we have imposed positivity of the eigenvalues through the Heaviside step functions $\theta$. The peaks are separated from each other in $\vec{r}$ and it is not easy to keep track such a point distribution. In order to overcome this technical difficulty, it is convenient to work with the homogeneous (independent of $\vec{r}$) average peak number density
\begin{align}
\begin{split}
\langle n_{\rm pk}(\vec{r}) \rangle 
&= \int \D{\nu} \D{x} \D{y} \D{z} ~P_{\rm pk}\left(\nu,x,y,z\right) \,,
    \label{eq:definition of number density}
\end{split}
\end{align}
where we have defined the {\it peak distribution function} as
\begin{align}
\begin{split}
P_{\rm pk}\left(\nu,x,y,z\right) 
&\equiv \frac{8\pi^{2}}{3} N y(y^{2}-z^{2})\chi(y,z)
\int \D^3\alpha \, n_{\rm pk}(\vec{r}) 
e^{-Q(\nu,\alpha,x,y,z)}
 \,.
\end{split}
\label{eq:pdf (peaks) of Gaussian}
\end{align}
Substituting $|{\rm det}\xi|=\sigma_2^3\lambda_1\lambda_2\lambda_3$ and $\delta^{(3)}(\vec{\eta})=\sigma_1^{-3} \delta^{(3)}(\vec{\alpha})$ in \eqref{n-peak} and then using the result together with \eqref{eq:definition of x,y,z} in \eqref{eq:pdf (peaks) of Gaussian} we find
\begin{align}
    \begin{split}
        P_{\rm pk}(\nu,x,e,p) =\frac{8\pi^{2}}{3^{4}} N \Big(\frac{\sigma_2}{\sigma_1}\Big)^3
        x^{8} e^{-Q_{\nu x}\left(\nu,x\right)} e^{-Q_{ep}\left(x,e,p\right)} \mathcal{J}(e,p) \Theta(x,e,p),
    \end{split}\label{eq:pdf (peaks) of Gaussian (ep)}
\end{align}
where, for the later convenience, we have decomposed function $Q\left(\nu,\alpha=0,x,y=ex,z=px\right)=Q\left(\nu,x,e,p\right)$ as $Q\left(\nu,x,e,p\right)=Q_{\nu x}\left(\nu,x\right) + Q_{ep}\left(x,e,p\right)$ with 
\begin{align}
\begin{split}
Q_{\nu x}\left(\nu,x\right) 
&\equiv \frac{\nu^{2} - 2\gamma \nu x + x^2}{2(1-\gamma^{2})} \, ,
\qquad
Q_{ep}\left(x,e,p\right) 
\equiv
\frac{5}{2} x^2 \left(3 e^2+p^2\right) \,,
\end{split}
\label{eq:Q(F,x) in Gaussian}
\end{align}
and we have also defined 
\begin{align}\label{eq:definition of J(e,p)}
\mathcal{J}(e,p) = e \left(e^2-p^2\right) (1-2 p) \big[(1+p)^2 -9 e^2 \big] \,,
\end{align}
and
\begin{equation}\label{eq:definition of chi(e,p)}
    \Theta(x,e,p) =
        \begin{cases}
        1 & \left(0\leq e\leq \frac{1}{4},\quad -e\leq p\leq e,\quad x>0\right) \\
        1 & \left(\frac{1}{4}\leq e\leq \frac{1}{2},\quad 3e-1\leq p\leq e,\quad x>0\right) \\
        0 & \text{otherwise} \,.
        \end{cases}
\end{equation}

\subsection{Most probable values of spherical parameters}\label{subsec-spherical-parameter}
The conditional peak distribution function for the sphericity parameters subjects to fixed values of $(\nu,x)$ is given by
\begin{align}
    P_{\rm pk}(e,p|\nu,x) \D{e}\D{p} = \frac{P_{\rm pk}(\nu,x,e,p)\D{\nu}\D{x}\D{e}\D{p}}{P_{\rm pk}(\nu,x)\D{\nu}\D{x}} \,,
    \label{eq:2 dim PDF for e,p}
\end{align}
where
\begin{align}
    \begin{split}
        P_{\rm pk}(\nu,x) = \int P_{\rm pk}(\nu,x,e,p) \D{e}\D{p} =\frac{8\pi^{2}}{3^{4}} N \Big(\frac{\sigma_2}{\sigma_1}\Big)^3 f(x)e^{-Q_{\nu x}\left(\nu,x\right)},
    \end{split}
    \label{P-Fx}
\end{align}
in which we have defined
\begin{align}
    \begin{split}
        f(x) \equiv x^{8}\int \D{e}\D{p} \mathcal{J}(e,p)\Theta(x,e,p)e^{-Q_{ep}\left(x,e,p\right) } \,.
    \end{split}
    \label{def-f}
\end{align}
The explicit form of $f(x)$ is given in Eq. \eqref{function-f}. Eq.~\eqref{eq:2 dim PDF for e,p} does not depend on $\nu$ because it only appears in the factors $e^{-Q_{\nu x}}$ defined in Eq.~\eqref{eq:Q(F,x) in Gaussian}, which do not involve $(e,p)$ and thus cancel out. Therefore, $P_{\rm pk}(e,p|\nu,x) = P_{\rm pk}(e,p|x)$ such that
\begin{align}
P_{\rm pk}(e,p|x) \D{e}\D{p} = \frac{x^8 e^{-Q_{ep}\left(x,e,p\right)} {\cal J}\left(e,p\right) \Theta\left(x,e,p\right) \D{x}\D{e}\D{p}}{f(x)\D{x}} \,.
\label{eq:2 dim PDF for e,p-final}
\end{align}

\begin{figure}
   \centering
    \includegraphics[width=0.5\linewidth]{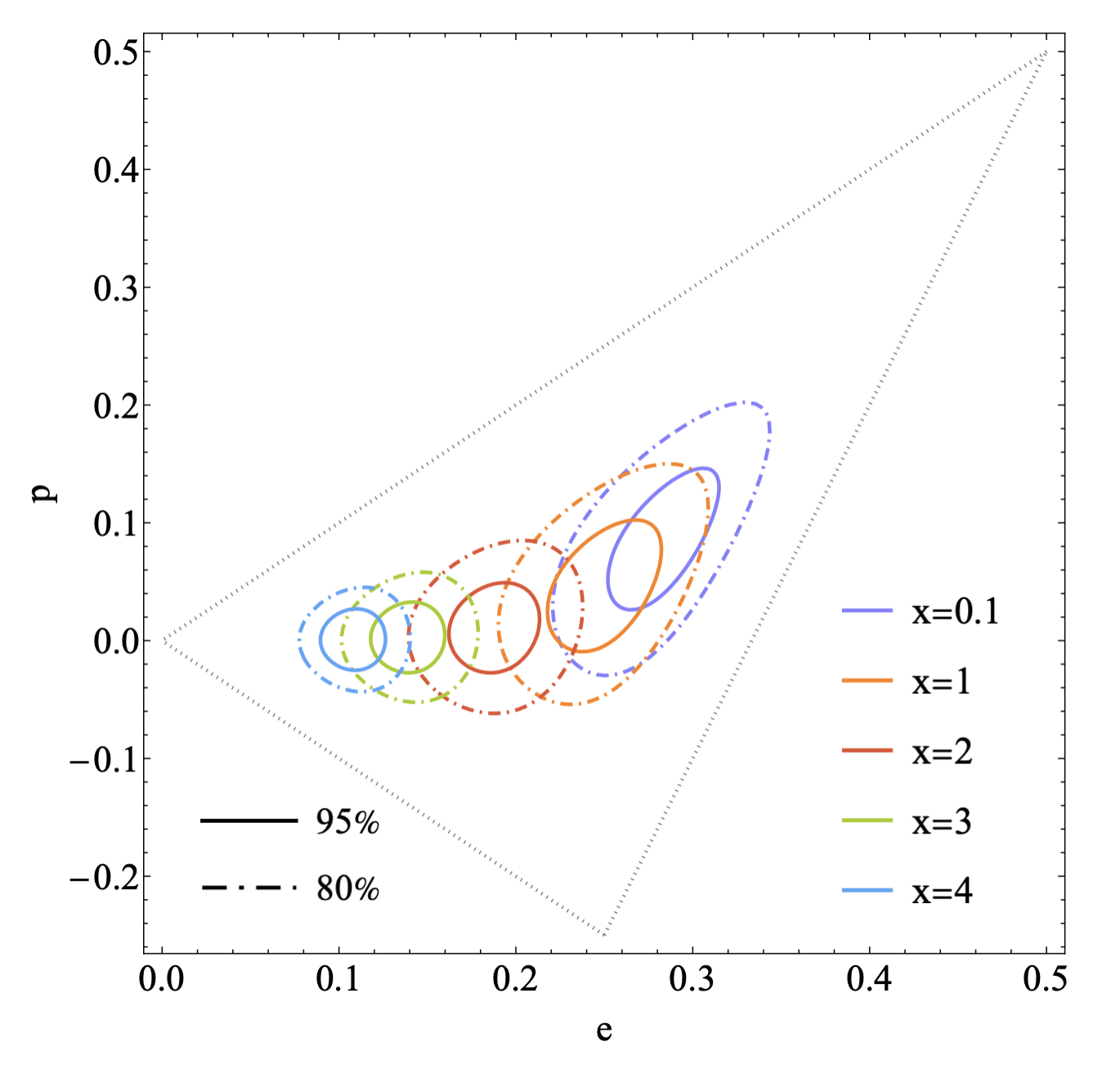}
  \caption{Contour plot of $P_{\rm pk}(e,p|x)$ defined in Eq.~\eqref{eq:2 dim PDF for e,p-final}. Peaks with large $x$ implies small $e$ and $|p|$. Moreover, we find $|p|\ll{e}$ in the regime of large $x$.}
    \label{fig:contourplot-g1}
 \end{figure}

Fig.~\ref{fig:contourplot-g1} shows the contour plot of $P_{\rm pk}(e,p|x)$ for fixed $x>0$. It is maximized at $E_m=\{e_m,p_m\}$ for a fixed value of $x$, that are subject to the condition $0 = \partial_{E}\ln P_{\rm pk}(e,p|x) = \partial_{E} \ln P_{\rm pk}(\nu,x,e,p)$ which gives 
\begin{align}
\big[ \partial_{E} Q_{ep}(x,e,p) - \partial_{E} \ln{\cal J}(e,p) \big]_{E=E_m} = 0 \,;
\qquad
E = \{ e, p \} \,.
\label{Eq-most-probable}
\end{align}

For large $x\gg 1$,\footnote{For $x\gg 1$, Eq.~\eqref{Eq-most-probable} is reduced to $x^{2}e=0$, i.e.~$e_{m}=0$. However, $P_{\rm pk}(e,p|x)$ also vanishes if $e$ vanishes. Thus, we consider large $x$ with finite $x^{2}e$ in our analysis, which implies $e\ll 1$.} as it can be clearly seen from Fig.~\ref{fig:contourplot-g1}, $E_{m}=\{e_{m},p_{m}\}$ satisfies $|p_{m}|\ll e_{m}\ll1$. In that region, we have
\begin{align}
\begin{split}
\partial_{e}\ln\mathcal{J}(e,p)
&\simeq 3e^{-1}-18e+ \mathcal{O}\left(e^{2},p^2e^{-3}\right) \,,
\\
\partial_{p}\ln\mathcal{J}(e,p)
&\simeq 18e^{2}-2pe^{-2} + \mathcal{O}\left(e^{3},p^3e^{-4},p\right) \,.
\end{split}
\end{align}
Substituting the above results in Eq. \eqref{Eq-most-probable}, we find the following  solutions for the sphericity parameters
\begin{align}
    e_{m}^{-2} \simeq  6+5x^{2}\simeq 5x^{2} ,
    \qquad
    p_{m} \simeq  6e_{m}^{4}\,,
    \qquad
    (\mbox{for a fixed large }\, x) \,. 
    \label{eq:em of Gaussian}
\end{align}
One should note that $E_{m}=\{e_{m},p_{m}\}$ is independent of the height of the peak $\nu$ but determined only by the value of large $x$, since peak distribution function Eq.~\eqref{eq:2 dim PDF for e,p-final} is independent of $\nu$. 
For $x\ll 1$, on the other hand, $E_{m}=\{e_{m},p_{m}\}\simeq \{0.28, 0.087\}$ becomes $x$-independent, because Eq.~\eqref{Eq-most-probable} is $x-$independent for $x^{2}e^{2}\ll 1$. 

To consider the asymptotic behavior of $E_{m}=\{e_{m},p_{m}\}$ for high peaks $\nu\gg1$, one should find $x=x_{m}$ for a fixed $\nu$ which maximizes $P_{\rm pk}(\nu,x)$ and substitute it into Eq.~\eqref{eq:em of Gaussian} to find $E_{m}=\{e_{m},p_{m}\}$ as a function of $\nu$. The value of $x_{m}$ is found by solving
\begin{align}
&\partial_x\ln{P_{\rm pk}(\nu,x)} = - \partial_x Q_{\nu x}(\nu,x) + \partial_x\ln{f}(x) = 0
\,,
\label{eq:dP/dchi-0}
\end{align}
where $Q_{\nu x}(\nu,x)$ and $f(x)$ are given by \eqref{eq:Q(F,x) in Gaussian} and \eqref{function-f}, respectively. In the asymptotic regions $x\gg1$ and $x\ll1$, as shown in \eqref{function-f-limits}, we have $f(x)\propto{x}^3$ and $f(x)\propto{x}^8$, respectively. Therefore, we have $\partial_x\ln{f}(x)=q/x$ where $q=3$ and $q=8$ for $x\gg1$ and $x\ll1$, respectively. Using \eqref{eq:Q(F,x) in Gaussian}, Eq. \eqref{eq:dP/dchi-0} gives $x_m \simeq  \left(\gamma\nu+\sqrt{4q(1-\gamma^2)+\gamma^2\nu^2}\right)/2$. If $\gamma=0$, $\nu$ does not correlate with $x$ and we obtain $x_{m}=\sqrt{q}$. In the case $\gamma\rightarrow 1$, $\nu$ and $x$ are highly correlated, thus we find $x_m\simeq\nu$ and $x_m\simeq0$ for $\nu\gg1$ and $\nu\ll-1$, respectively. We are not interested in the latter case with $x\ll1$ as $E_{m}=\{e_{m},p_{m}\}\simeq \{0.28, 0.087\}$ becomes $x$-independent. Therefore, let us focus on the regime $x\gg1$ and $\nu\gg1$ assuming $\gamma\neq 0$.
By using the asymptotic behavior of $f(x)$ for large $x$ is (see Eq. \eqref{function-f-limits}),
\begin{align}
    \begin{split}
        f(x)\simeq &\sqrt{\frac{2\pi}{5}}\frac{1}{3^{2}\cdot 5^{2}}\left( x^3 - 3x + \mathcal{O}\left(x^{-1}\right)\right)\quad({\rm large}\,x) \,,
        \label{eq:expression of f(x)}
    \end{split}
\end{align}
where the error is less than $1\,\%$ for $x>3$ \cite{Bardeen:1985tr}, we obtain
\begin{align}
    \begin{split}\label{xm}
        x_{m}\simeq \gamma\nu
        \left(1+\frac{3(1-\gamma^{2})}{\gamma^{2}\nu^{2}} \right)+ O\left(\frac{1}{\gamma^{2}\nu^{2}}\right)\,.
    \end{split}
\end{align}
Substituting it in Eq. \eqref{eq:em of Gaussian}, we find
\begin{align}
    \begin{split}
        e_{m}^{-2} \simeq  5\gamma^{2}\nu^{2} + 6 + 30(1-\gamma^{2})\,.
    \end{split}
\end{align}
Thus, $e_m\ll1$ for $\gamma\nu\gg1$. This is nothing but the well-known result of Ref. \cite{Bardeen:1985tr} that {\it higher peaks are more spherical}.

It is also useful to approximate Eq.~\eqref{eq:2 dim PDF for e,p-final} by a Gaussian function around the point $(e_{m},p_{m})$. Doing so, we find 
\begin{align}
    P_{\rm pk}(e,p|x)\simeq P_{\rm pk}(e_m,p_m|x)\exp\left(-\frac{(e-e_m)^{2}}{2\sigma_{ee}^{2}} -\frac{(e-e_m)(p-p_m)}{\sigma_{ep}^{2}} -\frac{(p-p_m)^{2}}{2\sigma_{pp}^{2}}\right)\,,
\end{align}\label{eq:pdf-ep-approx}
where the parameters $\sigma_{EE'}$ are interpreted as the variances if the approximation Eq.~\eqref{eq:pdf-ep-approx} is valid for $\sqrt{(E-E_{m})(E'-E'_{m})}\geq \sigma_{EE'}$. The variances obtained from
\begin{align}\label{def-sigmaAB}
\sigma_{EE'} ^{-2} 
&= |\partial_{E}\partial_{E'}\ln P_{\rm pk}(e,p|x) |
\\
&= | \partial_{E}\partial_{E'} Q_{e,p}\left(x,e,p\right) - \partial_{E}\partial_{E'}\ln{\cal J}\left(e,p\right) | \,,
\end{align}
where $E, E' = \{ e, p \}$ and we have used Eq.~\eqref{eq:2 dim PDF for e,p-final} in the second line. Now, using
\begin{align}
\begin{split}
\partial_{e}^2\ln\mathcal{J}(e,p)
&\simeq 
-3e^{-2} - 18 + \mathcal{O}\left(p,e^2,p^2e^{-4}\right) \,,
\\
\partial_{p}^2\ln\mathcal{J}(e,p)
&\simeq 
-2e^{-2} -6
+\mathcal{O}\left(p,e^2,p^2e^{-4}\right) \,,
\\
\partial_e\partial_{p}\ln\mathcal{J}(e,p)
&\simeq 
36 e + 4 p e^{-3}
+\mathcal{O}\left(ep\right) \,,
\end{split}
\label{J-derivatives-2nd}
\end{align}
we find
\begin{align}
    \sigma_{ee}^{-2}\simeq 6e_{m}^{-2},\quad 
    \sigma_{pp}^{-2} \simeq  3e_{m}^{-2},\quad  
    \sigma_{ep}^{-2} \simeq  60e_{m}.\label{eq:sigma(e,p)}
\end{align}
Note that $e_{m}/\sigma_{ee}\sim O(1)$ which implies that $P_{\rm pk}(e,p|x)$ is not suppressed for $0\lesssim e \lesssim 2e_{m}$ if the approximation Eq.~\eqref{eq:pdf-ep-approx} is valid in such a region.

\subsection{Troughs}\label{sec:valleys}
The analysis up to this point is valid for peaks with $x>0$ for both $\nu>0$ and $\nu<0$. However, generalization to the case of troughs with $x<0$ is quite straightforward as follows: The Gaussian PDF \eqref{eq:7 dim PDF of Gaussian (xyz)} is sensitive to the sign of $x$ only through the term $\nu x$ in the quadratic function $Q(\nu,\alpha,x,y,z)$ that is defined in Eq.~\eqref{eq:Q0 of Gaussian}. Thus, the configurations with $\{x\geq 0,\nu\geq 0\}$ and $\{x\leq 0,\nu\leq 0\}$ have the equivalent statistics, and similarly, those with $\{x\geq 0,\nu\leq 0\}$ and $\{x\leq 0,\nu\geq 0\}$ have the equivalent statistics. 
Therefore, one can easily construct the results for $x<0$ from the above results. For $x<0$, $\nu\gg 1$ induces $x_{m}\simeq 0$ thus $\{e_{m},p_{m}\}\simeq \{0.28, 0.087\}$, while $\nu\ll -1$ induces $x_{m}\simeq -\gamma|\nu|$ and we find $e_{m}^{-2} \simeq  5\gamma^{2}\nu^{2} + 6 + 30(1-\gamma^{2})$.

\section{Peaks with general local-type non-Gaussianity}\label{sec:exact local nonG}
There are many ways to go beyond the ideal Gaussian statistics discussed in the previous section. Among these, the so-called local non-Gaussianity deals with the case in which the statistical variable of the system under consideration is a nonlinear function of a Gaussian variable obeying a Gaussian statistics. 

The general form of local non-Gaussianity map is given by
\begin{align}
    F(\vec{r})= F\left[F_{G}(\vec{r})\right] \,,
    \label{eq:definition of local type F}
\end{align}
where $F$ is a general functional of $F_{G}$ without convolution in the $\vec{r}$ space. Note that in this section we label all Gaussian variables with the subscript $G$ while the non-Gaussian variables, like $F$, do not have any subscript.
When the map \eqref{eq:definition of local type F} is linear, of course, we deal with the Gaussian statistics that we studied in the previous section. Therefore, the non-Gaussian features show up through a nonlinear map between the variable of interest $F(\vec{r})$ and the Gaussian variable $F_{G}$.

To deal with such non-linearity, we shall first demonstrate the analysis of a statistical variable that is a function of a Gaussian random variable, $F=F(F_G)$, before considering its spatial shape, i.e.~$F(\vec{r})=F[F_{G}(\vec{r})]$.
Now, we require that $F$ is real valued in the full range $-\infty<F_{G}<\infty$ which is necessary for a Gaussian random variable.
For example, let us consider 
\begin{align}
F = F_{G} + f_{\rm NL}(F_{G}^{2}-\sigma_{G,0}^{2}) \,.
\label{eq:fNL expansion of F-L}
\end{align}
Here, $\sigma_{G,0}^2 \equiv \langle F_{G}^{2} \rangle$.
Independently on a non-zero constant $f_{\rm NL}$, $F$ is clearly defined in the full range of $F_G$.

One may now ask whether the PDF of $F$ can be rewritten in terms of a differential in $\D F_G$ rather than in terms of $\D F$. This is always possible, as long as one is careful to split the integral adequately around the zero of the Jacobian of the transformation $\D F\rightarrow \D F_G$. For example, the probability for $F$ to be in between $F_1$ and $F_2$ is given by
\begin{equation}
	\int_{F_1}^{F_2} P(F) \D F\ ,
\end{equation}
where $P(F)$ is the PDF of $F$. To obtain the expression in terms of the probability of $F_G$, we use the inversion,
\begin{equation}
	F_G^{\pm}=\frac{1}{2f_{\rm NL}}\left(-1\mp \sqrt{1+4 f_{\rm{NL}}\left(F+\sigma _{G,0}^2 f_{\rm{NL}}\right)} \right)\ .
\end{equation}
We see that there are two roots as the inversion is not univocal. This is not a problem. It simply telling us that the probability in the range $F_{1}\leq F\leq F_{2}$ is the sum of two probabilities in $F_G$ ranging appropriately. 
In other words,
\begin{eqnarray}
	\int_{F_1}^{F_2} P(F) \D F&=&\int_{F_1}^{F_2} \left|\frac{\D F_G^+}{\D F}\right| P(F_G^+) \D F+\int_{F_1}^{F_2} \left|\frac{\D F_G^-}{\D F}\right| P(F_G^-) \D F\\
	&=&\int_{F_G^+(F_1)}^{F_G^+(F_2)} P(F_G) \D F_G+\int_{F_G^-(F_1)}^{F_G^-(F_2)} P(F_G) \D F_G\ .\label{simpler}
\end{eqnarray}
Importantly, one should note that, $F$ might be bounded by its relation with $F_G$, even though $F_G$ is unbounded. For example, in Eq.~\eqref{eq:fNL expansion of F-L}, $F$ is bounded as $F\leq \frac{1}{4|f_{\rm NL}|}+\sigma _{G,0}^2|f_{\rm NL}|$ for $f_{\rm NL}<0$.

We are now ready to explore the shape dependence of $F$ as in Eq.~\eqref{eq:definition of local type F}. We have,
\begin{align}\begin{split}
&F_{G}=F_{G}(F),\\
&{\eta_{G}}_{i} = {\eta_{G}}_{i}\left(F,\eta\right) =J_{1}(F){\eta}_{i},\\
&{\xi_{G}}_{ij} = {\xi_{G}}_{ij}\left(F,\eta,\xi\right) = J_{1}(F){\xi}_{ij} + J_{2}(F){\eta}_{i}{\eta}_{j} \,,
\label{eq:hG of exact local nonG}
\end{split}
\end{align}
where we defined 
\begin{align}\label{Jacobian-local-NG}
J_{n}(F)=\frac{\delta^{n} F_{G}}{\delta F^{n}} \,,
\end{align}
in which a $\delta$ denotes the functional derivative. 

We know that $F_G$ obeys Gaussian statistics and, therefore, the inverse map \eqref{eq:hG of exact local nonG} allows us to find the statistics of the non-Gaussian random field $F$. In this regard, all we need to do is to substitute \eqref{eq:hG of exact local nonG} in the results of the last section. Let us emphasize again that $F_G, {\eta_{G}}_{i}, {\xi_{G}}_{ij}$ represent the Gaussian variables that we dealt in the previous section where we did not use the subscript ``$G$". 

Using the map \eqref{eq:hG of exact local nonG}, we see that the measure in the PDF \eqref{PDF} changes as
\begin{align}\label{measure-NG}
\D{F}_G\D^3\eta_G\D^6\xi_G = {J_{1}(F)}^{10} \D{F}\D^3\eta\D^6\xi \,,
\end{align}
where the Jacobian is
\begin{align}
\begin{split}
\det\left(\frac{\partial(F_{G},\eta_{G},\xi_{G})}{\partial(F,\eta,\xi)}\right) = {J_{1}(F)}^{10} \,.
\end{split}
\end{align}
Note that $J_2(F)$ does not contribute in the above determinant. The explicit form of $J_1(F)$ can be read for any given 
map \eqref{eq:definition of local type F} (as long as one is careful to treat separately the different branches of the inverse map $F_G\rightarrow F$). Starting from \eqref{PDF}, we find
\begin{align}\label{PDF-NG}
\begin{split}
P\left(F,\eta,\xi\right) {J_{1}(F)}^{10} \D F\D^3\eta \D^6\xi \propto&  \,e^{-Q\left(F,\eta,\xi\right)} {J_{1}(F)}^{10} \D F\D^3\eta \D^6\xi  \,,
\end{split}
\end{align}
where 
\begin{align}\label{PDF-NG-def}
P\left(F,\eta,\xi\right) &=
P_G\left[F_G=F_G(F), {\eta_{G}}_{i}\left(F,\eta\right), {\xi_{G}}_{ij}\left(F,\eta,\xi\right) \right] \,,
\\
\label{Q-NG-def}
Q\left(F,\eta,\xi\right) &= Q_G\left[F_G=F_G(F), {\eta_{G}}_{i}\left(F,\eta\right), {\xi_{G}}_{ij}\left(F,\eta,\xi\right) \right] \,.
\end{align}
Note that in \eqref{Q-NG-def}, we need to substitute the Gaussian variables $F_G, {\eta_{G}}_{i}, {\xi_{G}}_{ij}$ in terms of the non-Gaussian variables $F, {\eta}_{i}, {\xi}_{ij}$ through the inverse map \eqref{eq:hG of exact local nonG}. Thus, the functional form of $Q$ as a function of non-Gaussian variables $(F,\eta,\xi)$ is different from the functional form of $Q_G$ as a function of Gaussian variables $(F_G,\eta_G,\xi_G)$ as far as the map \eqref{eq:definition of local type F} is nonlinear. 

As already discussed, we can interchangeably work with either $F$ or $F_G$. 
Working in terms of $F_G$, the PDF \eqref{PDF-NG} takes exactly the same form as the Gaussian PDF \eqref{PDF}. However, this does not mean that we deal with Gaussian statistics of a physical field since the physical quantity of interest is $F$ and not $F_G$. Here, $F_G$ is a mathematical variable in terms of which the non-Gaussian PDF takes a Gaussian form: as long as the map \eqref{eq:definition of local type F} is nonlinear, we of course deal with a non-Gaussian statistics. 

We are interested in the statistics of the peaks where $\eta=0$. The inverse map \eqref{eq:hG of exact local nonG} takes a simpler form of ${\xi_{G}}_{ij} = {\xi_{G}}_{ij}\left(F,\xi\right) = J_{1}(F){\xi}_{ij}$ and we find
\begin{align}
{\Lambda_{G}}_{i} = J_{1}(F)\, {\Lambda}_{i} \,,
\end{align}
where ${\Lambda_{G}}_{i}$ and ${\Lambda}_{i}$ are the eigenvalues of the Gaussian $\left(-{\xi_{G}}_{ij}\right)$ and non-Gaussian $\left(-{\xi}_{ij}\right)$ variables, respectively. Note that ${\Lambda_{G}}_i=\sigma_{G,2}{\lambda_G}_i$ where ${\lambda_G}_i$ is the eigenvalues of the normalized variable $\left(-{\varsigma_{G}}_{ij}\right)=\left(-{\xi_{G}}_{ij}/\sigma_{G,2}\right)$ that is defined in \eqref{h-Normalized}. Taking into account the above observations, from \eqref{eq:definition of x,y,z} we find
\begin{align}\label{variables-change-NG}
X_G(F,X) &= J_1(F) X \,,
\qquad
e_G = e \,,
\qquad
p_G = p \,,
\end{align}
where $X_G = \sigma_{G,2} x_G$ and 
\begin{align}
X \equiv \sum_i \Lambda_i \,.
\end{align}
Now, using \eqref{measure-NG} and \eqref{variables-change-NG} in \eqref{eq:pdf (peaks) of Gaussian (ep)}, we find the peak distribution function,
\begin{align}
\begin{split}
P_{\rm pk}\left(F,X, e, p\right) \propto \frac{8\pi^{2}}{3^{4}}  \Big(\frac{\sigma_{G,2}}{\sigma_{G,1}}\Big)^3
X^{8} {J_1(F)}^{10} e^{-Q_{FX}\left(F,X\right)} e^{-Q_{ep}\left(F,X,e,p\right)} \mathcal{J}(e,p)\Theta(X_{G}(F,X),e,p),
\end{split}\label{eq:pdf (peaks) of Gaussian (ep)-NG}
\end{align}
with
\begin{align}
\begin{split}
Q_{FX}\left(F,X\right)
&\equiv \frac{1}{2(1-\gamma_G^{2})}\left(\frac{F_G(F)^{2}}{\sigma_{G,0}^{2}} - 2\gamma_G\frac{F_G(F)}{\sigma_{G,0}}\frac{X_{G}(F,X)}{\sigma_{G,2}}+\frac{X_{G}^{2}(F,X)}{\sigma_{G,2}^{2}}\right) ,
\\
Q_{ep}\left(F,X,e,p\right)
&\equiv
\frac{5 X_{G}^2(F,X)}{2 \sigma _{G,2}^2} \left(3 e^2+p^2\right) \,,
\end{split}
\label{eq:Q(F,x) in Gaussian-NG}
\end{align}
where we have used \eqref{Q-NG-def}.

We have set $\eta=0$ in our analysis above from the beginning as we are interested in the peaks. As it can be explicitly seen from \eqref{eq:Q(F,x) in Gaussian-NG} (remember $X=\sum_i\Lambda_i$), the dependency of $Q(F,\eta=0,\xi)$ on $\xi_{ij}$ is only through the eigenvalues of $\xi_{ij}$. In appendix \ref{app-local-NG}, we have shown that this is not the case for the general case with $\eta\neq0$. 
\subsection{Most probable values of spherical parameters}\label{subsec-spherical-parameter-NG}
The conditional peak distribution function for the sphericity parameters subject to the fixed values of $(F,X)$ is given by Eq.~\eqref{eq:2 dim PDF for e,p}. Using \eqref{eq:pdf (peaks) of Gaussian (ep)-NG} in \eqref{P-Fx}, we find
\begin{align}
\begin{split}
P_{\rm pk}(F,X) = \int P_{\rm pk}(F,X,e,p)\D{e}\D{p} \propto \frac{8\pi^{2}}{3^{4}} \Big(\frac{\sigma_{G,2}}{\sigma_{G,1}}\Big)^3 {J_{1}(F)}^{2} f\left[J_{1}(F)X\right] e^{-Q_{FX}\left[F,J_1(F) X\right]} \,,
\label{eq:pdf (peaks) of exact local nonG (Fx)}
\end{split}
\end{align}
in which $f(x)$ is defined in \eqref{def-f} and its explicit form is given by \eqref{function-f}. The extra Jacobian factor ${J_1(F)}^{10}$ cancels out from the numerator and the denominator in the conditional peak distribution function \eqref{eq:2 dim PDF for e,p} such that $P_{\rm pk}(e,p|F,X) = P_{\rm pk}(e,p|J_1(F)X)$. Therefore, the non-Gaussian effects in the measure does not play any role for fixed values of $F$. 

Taking into account the fact that $e=e_G$ and $p=p_G$, the whole effects of non-Gaussianity show up only through the combination $J_1(F)X=X_G$ in \eqref{eq:2 dim PDF for e,p-final}. This means that all analysis for $E_{m}=\{e_{m},p_{m}\}$ in Sec.~\ref{subsec-spherical-parameter} are directly applicable by substituting $F_G=F_G(F)$ and $X_G=J_1(F)X$. However, we need to be careful that the physical variable of interest is $X$ and not $X_G$, i.e., sharp peaks are defined by $X/\sigma_2\gg1$ not $X_G/\sigma_{G,2}\gg1$. To be concrete, in the following we consider the two cases of $X/\sigma_2\gg1$ and $X/\sigma_2\ll1$, in which we assume $J_1>0$, the extension to $J_1<0$ is straightforward.

For $X/\sigma_2\gg1$, depending on the form of $J_1(F)$, $X/\sigma_2\gg1$ might not necessarily imply  $X_G/\sigma_{G,2}\gg1$. For example, in principle, there can be an extreme case when $X/\sigma_2\gg1$ implies $X_G/\sigma_{G,2}\ll1$ for sufficiently small values of $J_1(F)\left(\sigma_{2}/\sigma_{G,2}\right)$. In that case, as we have shown in the previous section, we find $E_{m}=\{e_{m},p_{m}\}\simeq \{0.28, 0.087\}$ which is independent of $X_G/\sigma_{G,2}$. Therefore, peaks may get less spherical for large $X/\sigma_{2}$, which is a very interesting result. 
On the other hand, assuming that $X/\sigma_2\gg1$ implies  $X_G/\sigma_{G,2}\gg1$, which is the case at least for the small (perturbative) non-Gaussianity, the values $E_{m}=\{e_{m},p_{m}\}$ that maximize $P_{\rm pk}(e,p|X_{G})$ for large but finite values of $X_{G}=J_{1}(F)X$ can be directly read from \eqref{eq:em of Gaussian} as
\begin{align}
e_{m}^{-2} \simeq  6+5\frac{X_{G}^{2}}{\sigma_{G,2}^{2}}  = 6+5\frac{J_{1}^{2}(F)X^{2}}{\sigma_{G,2}^{2}} ,
\qquad p_{m} \simeq  6e_{m}^{4}\,,
\qquad (\mbox{for a fixed large }\,J_{1}(F)X/\sigma_{G,2}).\, \label{eq:em of exact local nonG with fixed Fx}
\end{align}
For the peaks with large  $F_{G} \gg \sigma_{G,0}$, we find
\begin{align}
        \left(\frac{X_{G}}{\sigma_{G,2}} \right)_{m}
        \simeq  \left(\gamma_{G}\frac{F_{G}(F)}{\sigma_{G,0}}\right)\left\{1+ \frac{3(1-\gamma_{G}^{2})\sigma_{G,0}}{\gamma_{G}F_{G}(F)}\right\}\,+\mathcal{O}\left(\frac{\sigma_{G,0}^{2}}{\gamma_{G}^{2}F_{G}^{2}(F)} \right),
\end{align}
from Eq.~\eqref{xm}, and\footnote{The standard deviations $\sigma_{EE'}$, defined in \eqref{def-sigmaAB}, also takes the same form as the Gaussian case, given in Eq.~\eqref{eq:sigma(e,p)}.}
\begin{align}
    e_{m}^{-2}\simeq 5 \left(\gamma_{G} \frac{F_{G}(F)}{\sigma_{G,0}}\right)^2 + 6 + 30(1-\gamma_{G}^{2}) \,.
\label{eq:em of exact local nonG with fixed F}
\end{align}

Now let us look at the case of $X/\sigma_{2}\ll 1$. In the extreme case that $X/\sigma_{2}\ll 1$ implies $X_G/\sigma_{G,2}\gg 1$, we can safely use \eqref{eq:em of exact local nonG with fixed F} which shows that the system is more spherical for 
$X/\sigma_{2}\ll 1$.
On the other hand, if $X/\sigma_{2}\ll 1$ implies $X_G/\sigma_{G,2}\ll 1$, which is the case at least for the small (perturbative) non-Gaussianity, we can use the result $E_{m}=\{e_{m},p_{m}\}\simeq \{0.28, 0.087\}$ as obtained in the previous section for the Gaussian case. We conclude that the shape in this case is much less spherical than the former case.

One should note that 
the nature of peaks is determined by the non-Gaussian variable $F$ and its derivatives, while the statistical properties such as the most probable values are determined by the Gaussian variable $F_{G}$ and its derivatives, 
which implies that rarer configurations tend to be more spherical. 
Moreover, the results \eqref{eq:em of exact local nonG with fixed Fx} and \eqref{eq:em of exact local nonG with fixed F} are applicable as far as $X_{G}/\sigma_{G,2}$ is large which, in principle, may not imply large $X/\sigma_{2}$. Therefore, large non-Gaussianity can violate the condition $X/\sigma_{2}\gg1 \leftrightarrow X_G/\sigma_{G,2}\gg 1$ (or $X/\sigma_{2}\ll 1 \leftrightarrow X_G/\sigma_{G,2}\ll 1$). In the case $X/\sigma_{2} \gg 1 \to X_G/\sigma_{G,2} \ll 1$, we find the interesting result that larger non-Gaussianity makes peaks less spherical, although we do not give any explicit examples of this for local-type non-Gaussianity.

\section{Non-Gaussian peaks: The role of bispectrum}
\label{sec:non-Gaussian peaks}

So far, our Universe seems to be extremely Gaussian distributed, at least in terms of the observed curvature perturbations. However, any model of inflation, which presumably generated those initial conditions, brings non-Gaussianities. While in average they might lead to small or even insignificant corrections to the Gaussian statistics, if we focus on rare events, non-Gaussianities might nevertheless dominate the Gaussian contribution. Although the effect on non-Gaussianities over the number of peaks has been already extensively studied (see e.g. \cite{Matsubara:2020lyv}), as far as we know, no attention has been given to the morphology of those peaks.  

To make our point, we formulate the statistics of peaks where a random field $F(\vec{r})$ is characterized only by the power spectrum defined in Eq.~\eqref{eq:definition of power-spectrum} and the bispectrum defined as
\begin{align}
\langle F(\vec{k}_{1})F(\vec{k}_{2})F(\vec{k}_{3})\rangle = (2\pi)^{3}\delta^{(3)}(\vec{k}_{1} + \vec{k}_{2} + \vec{k}_{3})B(k_{1},k_{2},\theta) \,,\quad \theta=\cos^{-1}{\left(\frac{\vec{k}_{1}\cdot\vec{k}_{2}}{k_{1}k_{2}}\right)} \,.
\label{bispectrum}
\end{align}
Here we have assumed homogeneity and isotropy such that the bispectrum $B(k_{1},k_{2},\theta)$ is invariant under the translation and rotation in the $\vec{k}$ space. In the perturbative regime, we neglect subleading contributions from higher-order correlators, such as trispectrum, so that the non-Gaussianity is fully characterized by the bispectrum. However, in general, higher-order correlators also contribute to the non-Gaussianities.

We first review the so-called Edgeworth expansion, which is a mathematical technique to obtain the PDF of random fields in the presence of a general type of non-Gaussianity in Sec.~\ref{subsec:general edgeworth}. We then study the effects of small non-Gaussianities on the peaks in Sec.~\ref{sec:general nonG}. We study large non-Gaussian effects in Sec.~\ref{sec-tail} by focusing on the tail of the PDF. 

\subsection{Review of Edgeworth expansion}\label{subsec:general edgeworth}

A sufficiently smooth general $d$-dimensional PDF $P(h)$ ($h=\{h_{1}\cdots h_{d}\}$) is completely determined by its set of cumulants $c_{n_1\ldots n_d}$. The statistical Fourier transformation of PDF, the characteristic function $z(k)$ ($k=\{k_{1}\cdots k_{d}\}$), is defined as
\begin{align}\label{eq:characteristic function}
\begin{split}
z(k)\equiv \int \D h e^{-ik^{T}h}P(h)=\langle e^{-ik^{T}h}\rangle = \exp{\left(\sum_{n_{1}=0}^{\infty}\cdots \sum_{n_{d}=0}^{\infty}\frac{c_{n_{1}\cdots n_{d}}}{n_{1}!\cdots n_{d}!}(-ik_{1})^{n_{1}}\cdots (-ik_{d})^{n_{d}}\right)}\,.
\end{split}
\end{align}
Equivalently, the cumulants can be found from $c_{n_{1}\cdots n_{d}}=\left.\frac{\partial^{n_{1}}}{\partial(-ik_{1})^{n_{1}}}\cdots \frac{\partial^{n_{d}}}{\partial(-ik_{d})^{n_{d}}}\ln z(k)\right|_{k=0}$. Cumulants have several important properties. First, they are related to the expected values $\langle h_{1}^{n_{1}}\cdots h_{d}^{n_{d}}\rangle $ as
\begin{align}
\begin{split}
\sum_{l=0}^{\infty}\frac{1}{l!}\left(\sum_{n_{1}=0}^{\infty}\cdots \sum_{n_{d}=0}^{\infty}\frac{c_{n_{1}\cdots n_{d}}}{n_{1}!\cdots n_{d}!}(-ik_{1})^{n_{1}}\cdots (-ik_{d})^{n_{d}}\right)^{l}=\sum_{n_{1}=0}^{\infty}\cdots \sum_{n_{d}=0}^{\infty}\frac{\langle h_{1}^{n_{1}}\cdots h_{d}^{n_{d}}\rangle}{n_{1}!\cdots n_{d}}(-ik_{1})^{n_{1}}\cdots (-ik_{d})^{n_{d}}\,,
\end{split}
\end{align}
which can be deduced from Eq.~\eqref{eq:characteristic function}. 
For the 1-dimensional case, it is reduced to
\begin{align}\begin{split}
c_{1} &= \langle h \rangle\,,
\\
c_{2} &= \langle h^{2} \rangle - \langle h \rangle^{2}\,,
\\
c_{3} &= \langle h^{3} \rangle -3 \langle h \rangle\langle h^{2} \rangle + 2\langle h\rangle ^{3}\,,\\
c_{4} &=\langle h^{4}\rangle- 4 \langle h \rangle \langle h^{3}\rangle - 3 \langle h^{2}\rangle^{2} + 12 \langle h \rangle ^{2}\langle h^{2}\rangle  -6\langle h\rangle ^{4} \,,\\
&\vdots
\end{split}
\label{cumulants-CF}
\end{align}
Note that the 0-th cumulant is $c_{0\cdots 0}=\ln z(0)=1$, regardless of the PDF or dimension. One can see that, up to the 3rd order, cumulants and expected values coincide when $c_{1}=0$. This persist in any dimensions. For instance, for a 1-dimensional Gaussian PDF $P(h)=\frac{1}{\sqrt{2\pi \sigma^{2}}}\exp\left(-\frac{(h-\mu)^{2}}{\sigma^{2}}\right)$, the characteristic function is $z(k)=\exp\left(-ik\mu + \frac{1}{2}(-ik)^{2}\sigma^{2}\right)$ and the cumulants are $c_{0}=0$, $c_{1} = \mu$, $c_{2} = \sigma^{2}$, $c_{n\geq 3}=0$ as expected from $\langle h\rangle =\mu$ and $\langle h^{2}\rangle  = \sigma^{2}+\mu^{2}$.

Now, let us reconstruct a PDF from a given set of the cumulants. Once all of the cumulants are found, $P(h)$ can be obtained through the inverse Fourier transformation of $z(k)$ as
\begin{align}
\begin{split}
P(h) = \int \frac{\D^{d}k}{(2\pi)^{d}}\exp{\left(ik^{T}h + \sum_{n_{1}=0}^{\infty}\cdots \sum_{n_{d}=0}^{\infty}\frac{c_{n_{1}\cdots n_{d}}}{n_{1}!\cdots n_{d}!}(-ik_{1})^{n_{1}}\cdots (-ik_{d})^{n_{d}}\right)}\,.
\end{split}
\end{align}
However, it is not an easy task to integrate over the $k$ space in general. Instead, we consider to expand it around the Gaussian PDF $P_{G}(h)$ whose cumulants $\gamma_{n_{1}\cdots n_{d}}$ are given by $\gamma_{n_{1}\cdots n_{d}}=c_{n_{1}\cdots n_{d}}$ for $\sum_{i=1}^{d}n_{i}\leq 2$ and $\gamma_{n_{1}\cdots n_{d}}=0$ for $\sum_{i=1}^{d}n_{i}>2$. Then, the PDF is reduced to
\begin{align}
\begin{split}
P(h)\propto
&\,\exp\left(\sum_{n_{1}=0}^{\infty}\cdots \sum_{n_{d}=0}^{\infty}\frac{c_{n_{1}\cdots n_{d}}-\gamma_{n_{1}\cdots n_{d}}}{n_{1}!\cdots n_{d}!}\left(-\partial _{h_{1}}\right)^{n_{1}}\cdots \left(-\partial_{h_{d}}\right)^{n_{d}}\right)P_{G}(h)\\
=  &\left(1+\sum_{n_{1}=0}^{\infty}\cdots \sum_{n_{d}=0}^{\infty}\frac{c_{n_{1}\cdots n_{d}}-\gamma_{n_{1}\cdots n_{d}}}{n_{1}!\cdots n_{d}!}\left(-\partial _{h_{1}}\right)^{n_{1}}\cdots \left(-\partial_{h_{d}}\right)^{n_{d}}+\cdots\right)P_{G}(h)\,,
\end{split}
\label{Edgeworth-expansion}
\end{align}
where the last dots in the parentheses include all cumulants. This is commonly known as Edgeworth expansion.\footnote{To work on the PDF perturbatively by truncating the expansions in Eq.~\eqref{Edgeworth-expansion}, the condition 
	\begin{align}
	\begin{split}
	\int \D^{d}{h} \left(\sum_{n_{1}=0}^{\infty}\cdots \sum_{n_{d}=0}^{\infty}\frac{c_{n_{1}\cdots n_{d}}-\gamma_{n_{1}\cdots n_{d}}}{n_{1}!\cdots n_{d}!}\left(-\partial _{h_{1}}\right)^{n_{1}}\cdots \left(-\partial_{h_{d}}\right)^{n_{d}}+\cdots\right)P_{G}(h)\ll \int \D^{d} h P_{G}(h)\,,
	\end{split}
	\end{align}
	should be satisfied, which is the case in the following section.	For instance, for the $1$-dimensional PDF with nonzero $c_{2}$ and $c_{3}$, we have
	$P(h) \propto\left(1 + \frac{c_{3}}{3!}(-\partial_{h})^{3}\right)\exp\left(-\frac{h^{2}}{2c_{2}}\right)
	=\left(1 + \frac{c_{3}}{3!\sqrt{c_{2}^{3}}}\left\{\frac{3h}{\sqrt{c_{2}}} -\frac{h^{3}}{\sqrt{c_{2}^{3}}}\right\}\right)\exp\left(-\frac{h^{2}}{2c_{2}}\right)$. This expansion is valid for $O(c_{3}h^{3}/c_{2}^{3})\ll1$.
	If there are small but non-vanishing higher cumulants such as $c_{4}$, we have to assume $c_{4}\ll O(c_{2}c_{3}/h)$.}

For our purpose, we need to consider $h=\{F,\eta,\xi\}$. In this case, the PDF takes the following form
\begin{align}\label{PDF-Edgeworth-gen}
P(F,\eta,\xi)\D{F}\D^3\eta \D^6\xi \propto \exp\left( \sum_{n=3}^{\infty} \frac{(-1)^n}{n!} 
\sum_{A_1,\cdots,A_n} 
c^{(n)}_{A_1 \cdots A_n} 
\partial_{h_{A_1}} \cdots \partial_{h_{A_n}} \right) {P}_{G}(F,\eta,\xi)\D{F}\D^3\eta \D^6\xi \,,
\end{align}
where
\begin{align}
{P}_{G}(F,\eta,\xi)\propto \exp{(-Q_G)} \,,
\qquad 
Q_G=\frac{1}{2}\sum_{A,B} h_A M^{-1}_{AB} h_{B} \,,
\label{def-QG}
\end{align}
are the same as the ones in the Gaussian statistics that we have discussed in Sec.~\ref{sec:Gaussian}. For the purpose of computation, it is useful to rewrite the PDF as follows
\begin{align}\label{PDF-Edgeworth-2}
P(F,\eta,\xi)\D{F}\D^3\eta \D^6\xi\propto K(F,\eta,\xi){P}_{G}(F,\eta,\xi)\D{F}\D^3\eta \D^6\xi \,,
\end{align}
where we have defined
\begin{align}\label{K-def-app-gen}
K(F,\eta,\xi) \equiv 
e^{Q_G} \exp\left( \sum_{n=3}^{\infty} \frac{(-1)^n}{n!} 
\sum_{A_1,\cdots,A_n} 
c^{(n)}_{A_1 \cdots A_n} 
\partial_{{A_1}} \cdots \partial_{{A_n}} \right)
e^{-Q_G} \,,
\end{align}
where $c^{(n)}_{A_1 \cdots A_n}$ are the cumulants and $\partial_{A_n}\equiv\partial/\partial{h_{A_n}}$. To compute $K(F,\eta,\xi)$, we have to operate the derivatives in the exponential to the right hand side and then simplify. As we will see soon, this is not an easy task in general. We thus look at different regimes to find simple expressions for $K(F,\eta,\xi)$.

\subsection{Small non-Gaussianity in peaks}\label{sec:general nonG}
In this subsection, we formulate the statistics of the peaks assuming that non-Gaussianity is small compared to Gaussian part. We thus assume that the 3-th cumulant is the dominant one among any non-Gaussian effects. Then, ignoring the 4-th and higher cumulants, \eqref{K-def-app-gen} is simplified to (see Appendix \ref{app-edgeworth} for the details of computations)
\begin{align}\label{K-final}
K(F,\eta,\xi) &\simeq 1 + \beta^{(1)}(F,\eta,\xi) + \beta^{(2)}(F,\xi) \,,
\end{align}
where
\begin{align}\label{def-beta}
\begin{split}
\beta^{(1)}(F,\eta,\xi) &= 
\frac{1}{3!} \beta_{ABC} M^{-1}_{AD} M^{-1}_{BE} M^{-1}_{CF} h_D h_E h_F \,,
\\
\beta^{(2)}(F,\xi) &=
- \frac{1}{2} \beta_{ABC} M^{-1}_{AD} M^{-1}_{BC} h_D
\,,
\end{split}
\end{align}
and 
\begin{align}
\beta_{ABC} \equiv c^{(3)}_{ABC} = \langle h_{A}h_{B}h_{C} \rangle \,,
\label{def-betas}
\end{align}
in which we have used the fact that the 3-th cumulant coincides with the 3-point function as shown in \eqref{cumulants-CF}. Thus, in the weak non-Gaussianity regime $\beta^{(1)}(F,\eta,\xi)\ll1$ and $\beta^{(2)}(F,\xi)\ll1$ characterize small deviation from the Gaussian case through the non-vanishing bispectrum \eqref{bispectrum}.

Now, our task is to find the explicit form of $\beta_{ABC}$, defined in \eqref{def-betas}, in terms of the bispectrum \eqref{bispectrum}. With direct computation, we find the following non-vanishing components of $\beta_{ABC}$ (see appendix \ref{app-I-ABC} for the details)
\begin{align}
\begin{split}
\beta_{FFF} 
&= {\cal I}_0 \,,
\qquad
\beta_{F\eta_i\eta_j} 
= - \frac{1}{2} \beta_{FF\xi_{ij}} 
= {\cal I}_1\, \delta_{ij} \,,
\\
\beta_{F\xi_{ij}\xi_{kl}} 
&= {\cal I}_2 \left( \delta_{ij}\delta_{kl} + \delta_{ik}\delta_{jl} + \delta_{il}\delta_{jk} \right) 
+ {\cal I}_3\, \delta_{ij}\delta_{kl} \,,
\\
\beta_{\eta_i\eta_{j}\xi_{kl}} 
&= \frac{1}{3}{\cal I}_3 \left(-2 \delta_{ij}\delta_{kl} + \delta_{ik}\delta_{jl} + \delta_{il}\delta_{jk} \right) \,,
\end{split}
\label{kappas}
\end{align}
and
\begin{align}
\begin{split}
\beta_{\xi_{ij}\xi_{kl}\xi_{mn}} 
&=
{\cal I}_4
\big[
3\delta_{ij} \delta_{kl} \delta_{mn} +\delta_{ij} (\delta_{kn} \delta_{lm} + \delta_{km} \delta_{ln})
+ (\delta_{in} \delta_{jm} + \delta_{im} \delta_{jn}) \delta_{kl} + (\delta_{il} \delta_{jk} + \delta_{ik} \delta_{jl}) \delta_{mn} \big]
\\
&+
{\cal I}_5 \big[
\delta_{il} (\delta_{jn} \delta_{km} + \delta_{jm} \delta_{kn}) 
+ (\delta_{ik} \delta_{jn} + \delta_{ij} \delta_{kn} ) \delta_{lm} + \delta_{in} (\delta_{jm} \delta_{kl} + \delta_{jl} \delta_{km} + \delta_{jk} \delta_{lm}) 
\\
&+ (\delta_{ik} \delta_{jm} + \delta_{ij} \delta_{km} ) \delta_{ln} + \delta_{im} (\delta_{jn} \delta_{kl} + \delta_{jl} \delta_{kn} + \delta_{jk} \delta_{ln}) + (\delta_{il} \delta_{jk} + \delta_{ik} \delta_{jl} + \delta_{ij} \delta_{kl}) \delta_{mn} 
\big] \,,
\end{split}
\label{kappa-xxx}
\end{align}
where we have defined
\begin{align}
{\cal I}_I \equiv \int \frac{\D^3{k}}{(2\pi)^3} \int \frac{\D^3{k'}}{(2\pi)^3}
B(k,k',\theta) I_I(k,k',\theta) \,;
\qquad
\theta = \cos^{-1}\bigg(\frac{{\vec k}\cdot{\vec k}'}{kk'}\bigg) \,.
\label{def-cal-I}
\end{align}
The explicit form of the integrands $I_I$ are given by
\begin{align}
    \begin{split}
        I_0 &= 1 \,,
        \quad 
        I_1 = \frac{1}{9} \left( k^2 + k'^2 + {\vec k}\cdot{\vec k}' \right) \,,
        \quad
        I_2 =
        \frac{1}{45} 
        \left[ \left(k^2+k'^2\right) |{\vec k}+{\vec k}'|^2
        +k^2k'^2
        \right]
        - \frac{1}{10} |{\vec k}\times{\vec k}'|^2
        \,,
        \\
        I_3 &= \frac{1}{6} |{\vec k}\times{\vec k}'|^2 \,,
        \quad
        I_4 =
        -\frac{1}{10} |{\vec k}\times{\vec k}'|^2I_{1}\,,
        \quad
        I_5 = -\frac{9}{70}
        ({\vec k}\cdot{\vec k}')^2 I_1 
        + \frac{1}{210} k^2 k'^2 \left( k^2+k'^2 - {\vec k}\cdot{\vec k}' \right)\,.
    \end{split}
    \label{functions-I}
\end{align}
For a given form of bispectrum $B(k,k',\theta)$, in principle, we can perform the integrations over the momenta in \eqref{def-cal-I} to find explicit forms of ${\cal I}_I$.

Having computed correlators $\beta_{ABC}$ in Eq. \eqref{kappas} and \eqref{kappa-xxx}, using \eqref{eq:2 pt correlation}, it is straightforward to compute $\beta^{(1,2)}$ given by \eqref{def-beta}. Similar to the previous section, we diagonalize $\xi_{ij}$. We find that the PDF depends not only on the eigenvalues of $\xi_{ij}$ but also on the three other independent components of $\xi_{ij}$ i.e. Euler angles $\vartheta_i$. Then, substituting \eqref{K-final} in Eq.~\eqref{PDF-Edgeworth-2}, the PDF takes the form (see appendix \ref{app:genral-NG} for the details)
\begin{align}\label{PDF-NG-lambda}
\begin{split}
&P \left(\nu,\alpha,\varsigma\right)\D \nu \D^3\alpha \D^6\varsigma 
= \frac{1}{6} N 
\big[1+\beta^{(1)}(\nu,\alpha,\lambda,\vartheta) + \beta^{(2)}(\nu,\lambda) \big] e^{-Q_G\left(\nu,\alpha,\lambda\right)} 
\\ &\hspace{5cm}
\times |(\lambda_{1}-\lambda_{2})(\lambda_{2}-\lambda_{3})(\lambda_{3}-\lambda_{1})| \D\nu \D^3\alpha \D\lambda_{1}\D\lambda_{2}\D\lambda_{3} \D^3\Omega_{S^3}(\vartheta)\,,
\end{split}
\end{align}
where $\D^3\Omega_{S^3}(\vartheta)$ is the volume element of a unit 3-sphere and the explicit forms of $\beta^{(1,2)}$ are found in \eqref{beta1-Trace} and \eqref{beta2-Trace} as
\begin{align}
\begin{split}
\beta^{(1)}(\nu,\alpha,\lambda,\vartheta) &= 
b_{\nu^3} \nu^3
- b_{\nu^2\varsigma} \nu^2 \sum_i\lambda_i
+ \nu
\Big[
b_{\nu\varsigma^2} \big(\sum_i\lambda_i\big)^2
+ b_{\nu\varsigma\varsigma} \sum_i\lambda_i^2
\Big]
\\
&- b_{\varsigma\varsigma\varsigma} \sum_i\lambda_i^3
- \sum_i\lambda_i
\Big[
b_{\varsigma\varsigma^2} \sum_i\lambda_i^2
+ b_{\varsigma^3} \big(\sum_i\lambda_i\big)^2
\Big]
\\
&+ \sum_i\alpha_i^2 \Big[ b_{\nu\alpha^2} \nu
- b_{\alpha^2\varsigma} \sum_i\lambda_i
\Big]
- b_{\alpha\alpha\varsigma}
\sum_i\lambda_i (R_{ik}\alpha_k)^2 \,,
\end{split}
\label{beta1-lambda}
\end{align}
and 
\begin{align}\label{beta2-lambda}
\beta^{(2)}(\nu,\lambda) &= 
b_\nu \nu - b_\varsigma  \sum_i\lambda_i \,.
\end{align}
The coefficients $b_{ABC}$ (or simply denoted as $b$) are defined in Eqs. \eqref{coeff-c}, \eqref{coeff-c-beta-2}, \eqref{coeff-c-alpha} which depend on $\gamma$ and ${\cal I}_I$ that are defined in \eqref{kappas} and \eqref{kappa-xxx}.

We set $\alpha=0$ to estimate the number density of peaks. In this case, the integrand in \eqref{PDF-NG-lambda} becomes independent of $\vartheta_{i}$ and we can simply integrate over the volume of the 3-sphere as $\int\D^3\Omega_{S^3}(\vartheta)=2\pi^2$. Working with coordinates $(x,e,p)$ that are defined in Eq. \eqref{eq:definition of x,y,z} and considering the number density as \eqref{eq:definition of number density}, it is straightforward to find the peak distribution function defined in \eqref{eq:pdf (peaks) of Gaussian (ep)} as
\begin{align}
\begin{split}
P_{\rm pk}(\nu,x,e,p) &=
\big[1 + \beta^{(1)}_{\nu{x}}(\nu,x) + \beta^{(2)}_{\nu{x}}(\nu,x) + \beta^{(1)}_{ep}(\nu,x,e,p)  \big] P_{G,{\rm pk}}(\nu,x,e,p) \,,
\end{split}\label{eq:pdf (peaks) of NG (ep)}
\end{align}
where $P_{G,{\rm pk}}(\nu,x,e,p)$ is the Gaussian peak distribution function that is defined in \eqref{eq:pdf (peaks) of Gaussian (ep)},
\begin{align}\label{beta2-x}
\beta^{(2)}(\nu,x) &= 
b_\nu \nu - b_\varsigma x \,,
\end{align}
and for the later convenience, we have decomposed $\beta^{(1)}\left(\nu,\alpha=0,x,e,p,\vartheta\right)=\beta^{(1)}\left(\nu,x,e,p\right)$ as $\beta^{(1)}\left(\nu,x,e,p\right)=\beta^{(1)}_{\nu x}\left(\nu,x\right) + \beta^{(1)}_{ep}\left(\nu,x,e,p\right)$ with 
\begin{align}
\begin{split}
\beta^{(1)}_{\nu x}(\nu,x) &= b_{\nu^3} \nu^3 
- b_{\nu^2 \varsigma} \nu^2 x
+
\frac{1}{3} \left(3 b_{\nu\varsigma^2} + b_{\nu\varsigma\varsigma}\right) \nu x^2 
- \frac{1}{9} \left(9 b_{\varsigma^3}+3 b_{\varsigma\varsigma^2} + b_{\varsigma\varsigma\varsigma}\right) x^3 \,,
\\
\beta^{(1)}_{ep}(\nu,x,e,p) &=
\frac{2}{3} \left( b_{\nu\varsigma\varsigma} \nu - b_{\varsigma\varsigma^2} x \right) x^2 \left(3 e^2+p^2\right)
- \frac{2}{9} b_{\varsigma\varsigma\varsigma} x^3
\big[ 9 (1+p) e^2 + (3-p) p^2
\big] \,.
\end{split}
\label{def-beta-ep}
\end{align}

\subsubsection{Most probable values of spherical parameters}\label{subsec-spherical-parameter-NG-general}
The conditional peak distribution function for the sphericity parameters subject to the fixed values of $(\nu,x)$, is described by Eq.~\eqref{eq:2 dim PDF for e,p} in which $P_{\rm pk}(\nu,x,e,p)$ is specified in \eqref{eq:pdf (peaks) of NG (ep)}, and $P_{\rm pk}(\nu,x)$ is given by
\begin{align}
\begin{split}
P_{\rm pk}(\nu,x) = \int P_{\rm pk}(\nu,x,e,p) \D{e}\D{p} =\frac{8\pi^{2}}{3^{4}} N \Big(\frac{\sigma_{2}}{\sigma_{1}}\Big)^3 f_{\rm tot}(\nu,x)e^{-Q_{\nu x}\left(\nu,x\right)},
\end{split}
\label{P-Fx-NG}
\end{align}
where
\begin{align}
\begin{split}
f_{\rm tot}(\nu,x) &\equiv  
\big[ 1 + \beta^{(1)}_{\nu x}(\nu,x) + \beta^{(2)}(\nu,x) \big] f(x) + g_{\beta}(\nu,x) \,,
\end{split}
\label{def-f-NG}
\end{align}
with $f(x)$ is defined in \eqref{def-f} and 
\begin{align}\label{def-g}
g_{\beta}(\nu,x) \equiv x^{8}\int \D{e}\D{p} \mathcal{J}(e,p)\Theta(e,p) \beta^{(1)}_{ep}(\nu,x,e,p) e^{-Q_{ep}\left(x,e,p\right) }  \,.
\end{align}
The explicit forms of $f(x)$ and $g_{\beta}(\nu,x)$ are shown in Eqs. \eqref{function-f} and \eqref{function-g}, respectively.

Now we consider $E_{m}=\{e_{m},p_{m}\}$ that maximize $P_{\rm pk}(e,p|\nu,x)$ for fixed $(\nu,x)$. The $\nu$ dependence of $P_{\rm pk}(e,p|\nu,x)$ appears through subdominant non-Gaussian effect of the order of $O(b\nu^{3},b\nu)$. Therefore, based on the results for the Gaussian statistics that is shown in Fig.~\ref{fig:contourplot-g1}, we should still have $|p_{m}|\ll e_{m}\ll 1$ for $x\gg 1$. Using the fact that $\partial_{E}\ln P_{\rm pk}(\nu,x,e,p)\approx\partial_{E}\ln P_{G,{\rm pk}}(\nu,x,e,p)+\partial_{E}\beta^{(1)}_{ep}(\nu,x,e,p)=0$, it is straightforward to find the following results for $0\leq |p|\ll e\ll 1$,
\begin{align}\label{eq:em,pm general}
    \begin{split}
        e_{m}^{-2}&\simeq 6+5x^{2}+\frac{4}{3}[(b_{\varsigma\varsigma\varsigma} + b_{\varsigma\varsigma^{2}})x-b_{\nu\varsigma\varsigma}\nu] x^{2} \,,\\
        &= 6+5x^{2}-\frac{75}{1-\gamma^{2}} [({\tilde{\cal I}}_{2} + 3{\tilde{\cal I}}_{4}\gamma + 7{\tilde{\cal I}}_{5}\gamma )\nu -({\tilde{\cal I}}_{2}\gamma + 3{\tilde{\cal I}}_{4} + 7{\tilde{\cal I}}_{5} )x ] x^{2} \,,\\
        p_{m}&\simeq 6\left(1-\frac{1}{9}b_{\varsigma\varsigma\varsigma} x^{3} \right)e^{4}_{m} =6\left(1-\frac{125}{2}\tilde{\cal{I}}_{5} x^{3} \right)e_{m}^{4} \,,\\
        &(\mbox{for a fixed }\nu\mbox{ and a large } x),
    \end{split}
\end{align}
where we have defined normalized dimensionless quantities
\begin{align}
\tilde{\mathcal{I}}_0 &\equiv \frac{{\cal I}_0}{\sigma_0^3} \,,
\qquad
\tilde{\mathcal{I}}_1
\equiv \frac{{\cal I}_1}{\sigma_0\sigma_1^2} \,,
\qquad
\tilde{\mathcal{I}}_2
\equiv \frac{{\cal I}_2}{\sigma_0\sigma_2^2} \,,
\qquad
\tilde{\mathcal{I}}_3
\equiv \frac{{\cal I}_3}{\sigma_1^2\sigma_2} \,,
\qquad
\tilde{\mathcal{I}}_4
\equiv \frac{{\cal I}_4}{\sigma_2^3} \,,
\qquad
\tilde{\mathcal{I}}_5
\equiv \frac{{\cal I}_5}{\sigma_2^3} \,,
\label{def-I-N}
\end{align}
which are compatible with the normalized variables \eqref{h-Normalized}. Note that $b{\tilde h}^{3}\ll1$ (${\tilde h}_A = h_A/\sigma_{A} = \{ \nu, \alpha, \varsigma \} $) or ${\tilde{\cal I}}h^{3}\ll1$ guarantees that the non-Gaussianity corrections are small compared with the leading Gaussian parts.

Among the three standard deviation $\sigma_{EE'}$ defined in Eq.~\eqref{def-sigmaAB}, only $\sigma_{ep}$ is modified as
\begin{align}
    \begin{split}
        \sigma_{ep}^{-2} \simeq  40\left(1-\frac{1}{9}b_{\varsigma\varsigma\varsigma} x^{3} \right)e_{m}=40\left(1-\frac{125}{2}\tilde{\cal{I}}_{5}x^{3}\right)e_{m}\,,
    \end{split}
\end{align}
and the other two $\sigma_{ee}$ and $\sigma_{pp}$ take the same form as the Gaussian case shown in Eq.~\eqref{eq:sigma(e,p)}.

Finally we derive the asymptotic behavior of $E_{m}=\{e_{m},p_{m}\}$ for a large fixed height $\nu\gg 1$. In order to do so, we have to find the most probable value of $x$. By using the large argument expansion for $f(x)$ and $g_{\beta}(\nu,x)$ given in Eq.~\eqref{function-f-limits} and \eqref{asymptotic-g}, respectively, we find
\begin{align}
    \begin{split}
        x_{m} \simeq \gamma \nu \left(1+\frac{3(1-\gamma^{2})}{\gamma^{2}\nu^{2}} + b_{\rm NG}\nu\right)\,,
    \end{split}\label{xm-NG}
\end{align}
for $\nu\gg 1$, where
\begin{align}
    b_{\rm NG} \equiv 
    - \left( 1 - \gamma^2 \right) \left[
    \frac{1}{3}\gamma(9b_{\varsigma^{3}}+3b_{\varsigma\varsigma^{2}} + b_{\varsigma\varsigma\varsigma})
    - \frac{2}{3} (3b_{\nu\varsigma^{2}}+b_{\nu\varsigma\varsigma} ) + \frac{1}{\gamma} b_{{\nu^{2}}\varsigma} \right] 
    = - \frac{1}{2} \left({\tilde{\cal I}}_0 - 6 {\tilde{\cal I}}_1\right)
    \,.
\end{align}
Consistently, we find that {\it the non-Gaussian correction $b_{\rm NG} \nu$ never dominates over the subleading Gaussian correction
$O((1-\gamma^{2})/\gamma^{2}\nu^{2})$} since we are in the parameter regime of $O(b\gamma^{2}\nu^{3})\leq O(b\nu^{3})\ll 1$.

Substituting Eq.~\eqref{xm-NG} into Eq.~\eqref{eq:em,pm general} yields
\begin{align}\label{eq:asymptotic e,p for general nonG}
    \begin{split}
        e_{m}^{-2}&\simeq 
        5\gamma^{2}\nu^{2} + 6+30(1-\gamma^{2})-(\tilde{\cal{I}}_{0}-6\tilde{\cal{I}}_{1} + 15\tilde{\cal{I}}_{2})\gamma^{2}\nu^{3}\,,\\
        p_{m}&\simeq 
        6\left(1-\frac{125}{2}\gamma^{3} \tilde{\cal{I}}_{5}\nu^{3}\right)e_{m}^{4} \,,
    \end{split}
\end{align}
where we have substituted the explicit values of coefficients $b$ that are given in \eqref{coeff-c}.

For validation and consistency checking, in Appendix~\ref{sec:specific bispectrums}, we apply the formalism developed in Sec.~\ref{sec:exact local nonG} and Sec.~\ref{sec:general nonG} to the local-type non-Gaussianity given by Eq.~\eqref{eq:fNL expansion of F-L}. We find the explicit forms of the sphericity parameters and confirm that both setups yield the same results in the regime of small (perturbative) non-Gaussianity.

\subsection{Rare non-Gaussian peaks: Exponential tail}\label{sec-tail}

The results of Sec.~\ref{sec:general nonG} can be only applied to the case of small perturbative non-Gaussianity. More specifically, the form \eqref{K-final} for $K(F,\eta,\xi)$ in PDF is only valid for $\beta^{(1,2)}\ll1$. However, in the case of PBH for example, one focuses on very rare peaks where non-Gaussianities might be relevant and possibly large.

In general, dealing with large non-Gaussianity is not an easy task. Even if the results of Sec.~\ref{sec:exact local nonG} can be, in principle, used for large non-Gaussianity, dealing with the associated difficulties is quite subtle. In this section, we shall focus on the features led by the bispectrum as before, to overcome this difficulty.

In general, $K(F,\eta,\xi)$ has the non-perturbative form given by Eq.~\eqref{K-def-app-gen}. We need to operate derivatives $\partial_{A}$ on the Gaussian factor $Q_G$, that is defined in \eqref{def-QG} as $Q_G=\frac{1}{2}\sum_{A,B} h_A M^{-1}_{AB} h_{B}$. We immediately find
\begin{align}
\partial_{A} Q_G = M^{-1}_{AB} h_B \,,
\qquad
\partial_{A} \partial_{B} Q_G = M^{-1}_{AB} \,,
\qquad 
\partial_{A} \partial_{B} \partial_{C} Q_G = 0 \,,
\label{D-Q}
\end{align}
where the summation rule over the dummy indices is considered. We see that $\partial_{A} Q_G$ is linear in $h_A$ while $\partial_{A} \partial_{B} Q_G$ is independent of $h_A$. This simple observation shows that if we look at the tail
\begin{align}
h_A \gg \sigma_A \, ;
\qquad
\sigma_A = \{\sigma_F,\sigma_\eta,\sigma_\xi\} \,,
\end{align}
the terms which include the highest power in $\partial_{A} Q_G = M^{-1}_{AB} h_B$ dominate over all other terms. Taking this fact into account, it is straightforward to find 
\begin{align}
\begin{split}
K(F,\eta,\xi)\Big{|}_{h_A \gg \sigma_A} &\approx 1 
+ \frac{1}{3!} c^{(3)}_{ABC} \partial_{A}Q_G \partial_{B}Q_G \partial_{C}Q_G
+ \frac{1}{4!} c^{(4)}_{ABCD} \partial_{A}Q_G \partial_{B}Q_G \partial_{C}Q_G\partial_{D}Q_G + \cdots
\\
&
+ \frac{1}{2!} 
\left[
\Big( \frac{1}{3!} c^{(3)}_{ABC} \partial_{A}Q_G \partial_{B}Q_G \partial_{C}Q_G \Big)^2
+
\Big( \frac{1}{4!} c^{(4)}_{ABCD} \partial_{A}Q_G \partial_{B}Q_G \partial_{C}Q_G\partial_{D}Q_G \Big)^2 + \cdots
\right]
\\
&+ \cdots \,.
\end{split}
\nonumber
\end{align}
Using \eqref{D-Q} and then resume all terms, we find the following exponential form for the tail
\begin{align}
\begin{split}
K(F,\eta,\xi) \simeq e^{-Q_{NG}(F,\eta,\xi)} \,;
\qquad
h_A \gg \sigma_A \,,
\end{split}
\label{K-exp}
\end{align} 
where we have defined
\begin{align}
\begin{split}
Q_{NG}(F,\eta,\xi) \equiv -\sum_{n=3}^{\infty} \frac{1}{n!} 
	c^{(n)}_{A_1 \cdots A_n} 
	(M^{-1}h)_{A_1} \cdots (M^{-1}h)_{A_n} \,,
\end{split}
\label{Q-NG}
\end{align} 
in which $(M^{-1}h)_A=M^{-1}_{AB} h_B$. Ignoring the $4$-th and higher cumulants $c^{(3)}_{ABC}=\beta_{ABC}$, $c^{(n\geq 4)}_{AB\cdots}=0$, general result \eqref{K-exp} reduces to Eq.~\eqref{K-final-exp} as expected.

Substituting \eqref{K-exp} in \eqref{PDF-Edgeworth-2}, we find
\begin{align}
P(F,\eta,\xi) \D F\D^3\eta \D^6\xi \propto
e^{-Q_G\left(F,\eta,\xi\right) - Q_{NG}\left(F,\eta,\xi\right) }\D F\D^3\eta \D^6\xi  \,;
\qquad
h_A \gg \sigma_A \,.
\label{PDF-NG-exp}
\end{align}
The term with the highest order of cumulant always dominates $Q_{\rm NG}$ in the regime $|h_{A}/\sigma_{A}|\rightarrow \infty$. Thus, the PDF is normalizable only if one truncates the summation in Eq.~\eqref{Q-NG} up to an even order of cumulants. If one truncates up to an odd order, $Q_{\rm NG}$ is governed by an odd power of $h_{A}/\sigma_{A}$, which diverges at either of positive or negative infinity.

\begin{figure}
    \centering
    \includegraphics[width=0.5\linewidth]{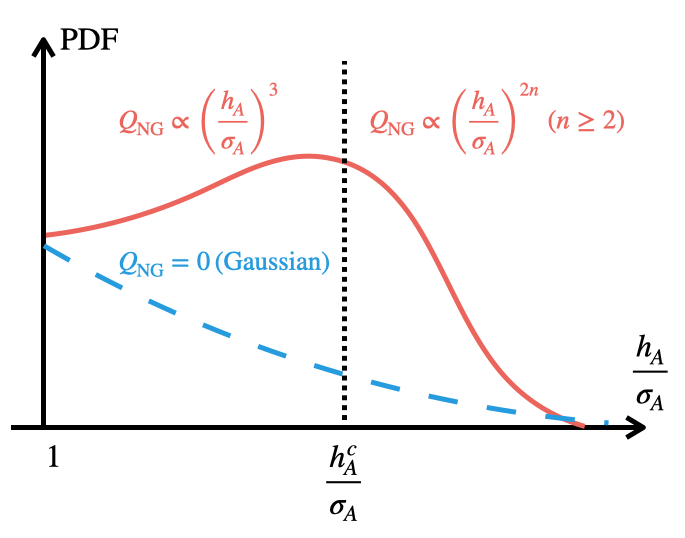}
    \caption{Schematic diagram of a PDF with large non-Gaussianity (red solid curve) in comparison with the perfect Gaussian PDF (blue dashed curve). The resummation of exponential in Eq.~\eqref{K-exp} is only possible for $h_{A}\gg \sigma_{A}$ and the normalization of the PDF is only possible if an even order of cumulant dominates for $h_{A}/\sigma_{A}\rightarrow \infty$.}
    \label{fig:schematic-pdf}
\end{figure}

In Eq.~\eqref{Q-NG}, the contributions of higher-order cumulants become increasingly significant as $h_A$
grows substantially larger than $\sigma_A$. More precisely, around $h_A\sim\sigma_A$, the non-Gaussian cumulants $c^{(n\geq 3)}_{AB\cdots}$ begin to dominate over the Gaussian term. The first cumulant to contribute is $c^{(3)}_{ABC}$, followed by $c^{(4)}_{ABCD}$ and so forth. Thus, as shown in Fig.~\ref{fig:schematic-pdf}, there will be a regime $1\ll h_{A}/\sigma_{A}\ll {h}^c_{A}/\sigma_{A}$ when the third cumulant $c^{(3)}_{ABC}$ completely dominates. Here $h^c_A$ corresponds to the value at which $c^{(4)}_{ABCD}$ becomes comparable to $c^{(3)}_{ABC}$. Note that the presence of a higher cumulant with an even order, such as $c^{(4)}_{ABCD}$ is essential to ensure the normalizability of the PDF at ${h}^c_{A}/\sigma_{A}\ll h_{A}/\sigma_{A}$. In this regard, in the regime $1\ll h_{A}/\sigma_{A}\ll {h}^c_{A}/\sigma_{A}$, \eqref{Q-NG} simplifies to
\begin{align}
    \begin{split}
        Q_{\rm NG}(F,\eta,\xi)
        &=-\beta^{(1)}(F,\eta,\xi)\,,
        \qquad 
        1\ll\frac{h_{A}}{\sigma_{A}}\ll \frac{{h}^c_{A}}{\sigma_{A}}\,,
    \end{split}
\label{eq:Q-NG-beta}
\end{align}
where $\beta^{(1)}$ is given by Eq.~\eqref{def-beta}. In
Fig.~\ref{fig:schematic-pdf}, we schematically compare the perfect Gaussian PDF $Q_{\rm NG}=0$ (blue dashed curve) with the non-Gaussian case $Q_{\rm NG}=-\beta^{(1)}\neq0$ (red solid curve) in the region $h_{A}>\sigma_{A}$.

We emphasize that the resummation to the exponential in Eq.~\eqref{K-exp} is only possible for $h_{A}\gg \sigma_{A}$ and the normalization of the PDF is only possible if an even order of cumulant dominates in the regime $h_{A}/\sigma_{A}\rightarrow \infty$. Under these assumptions, we can follow the same analysis as in Sec.~\ref{sec:general nonG} to find $P_{\rm pk}(e,p|\nu,x)$ while relaxing the assumption of $O(bh_{A}^{3}/\sigma_{A}^{3})\ll 1$. 
Namely, we obtain
\begin{align}
    \begin{split}
        P_{\rm pk}(\nu,x,e,p) &=
        e^{\beta^{(1)}_{\nu{x}}(\nu,x) + \beta^{(1)}_{ep}(\nu,x,e,p)  } P_{G,{\rm pk}}(\nu,x,e,p) \,,
        \qquad 1\ll \nu,x\ll {\nu}^c,{x}^c\,,
    \end{split}
\end{align}
where ${\nu}^c,{x}^c$ correspond to the region where higher cumulants dominate $Q_{\rm NG}$. Using the above result in \eqref{eq:2 dim PDF for e,p} we find 
\begin{align}
P_{\rm pk}(e,p|\nu,x) \D{e}\D{p} = \frac{e^{\beta_{\nu x}^{(1)}(\nu,x)+\beta^{(1)}_{ep}(\nu,x,e,p)  } P_{G,{\rm pk}}(\nu,x,e,p)\D{\nu}\D{x}\D{e}\D{p}}{P_{\rm pk}(\nu,x)\D{\nu}\D{x}} \,,
\label{eq:2 dim PDF for e,p-exp}
\end{align}
where
\begin{align}
P_{\rm pk}(\nu,x) = e^{\beta_{\nu x}^{(1)}(\nu,x)}\int \D e\D p\, e^{\beta^{(1)}_{ ep}(\nu,x,e,p)}{P_{G}}_{,\rm pk}(\nu,x,e,p) \,.
\end{align}
One can find $E_{m}=\{e_{m},p_{m}\}$ by solving $\partial_{E}P_{\rm pk}(e,p|\nu,x)=0$, which does not depend on $\beta_{\nu x}^{(1)}$.
Note that only $b_{\nu\zeta\zeta},b_{\zeta\zeta\zeta}$ and $b_{\zeta\zeta^{2}} $ appears in $\beta^{(1)}_{ ep}$ \eqref{def-beta-ep} and, thus, other $b$ parameters are irrelevant to determine the sphericity of peaks. 

\begin{figure}
    \centering
    \includegraphics[width=0.5\linewidth]{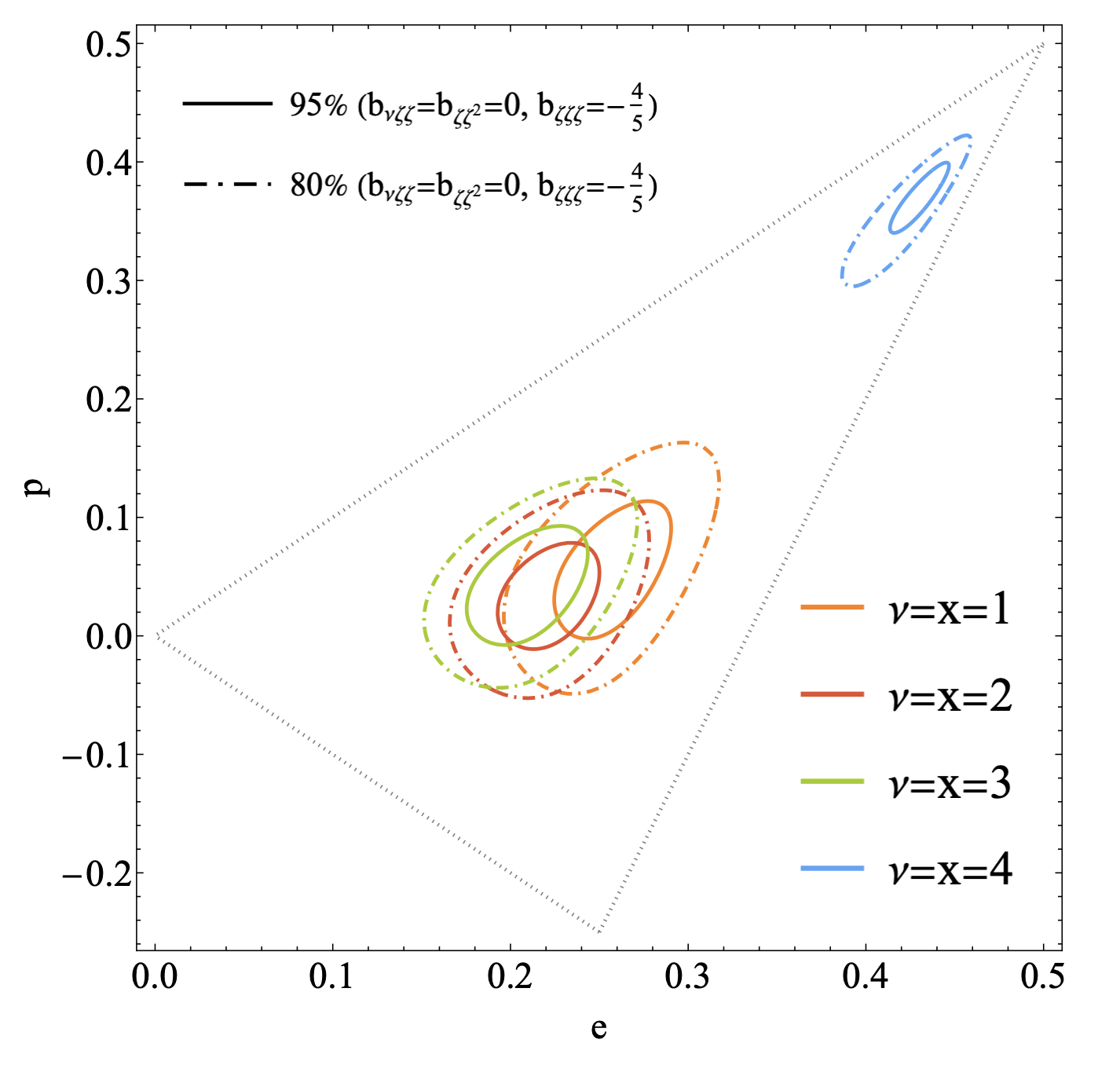}
    \caption{Contour plot of $P_{\rm pk}(e,p|\nu,x)$ for fixed $\nu$ and $x$ with $b_{\nu\zeta\zeta}=b_{\zeta\zeta^{2}}=0\,,\,b_{\zeta\zeta\zeta} = -4/5$. In this particular example, large non-Gaussianity makes the higher peaks less spherical.}
    \label{fig:contourplot-ng}
\end{figure}

\begin{figure}
    \centering
    \includegraphics[width=0.5\linewidth]{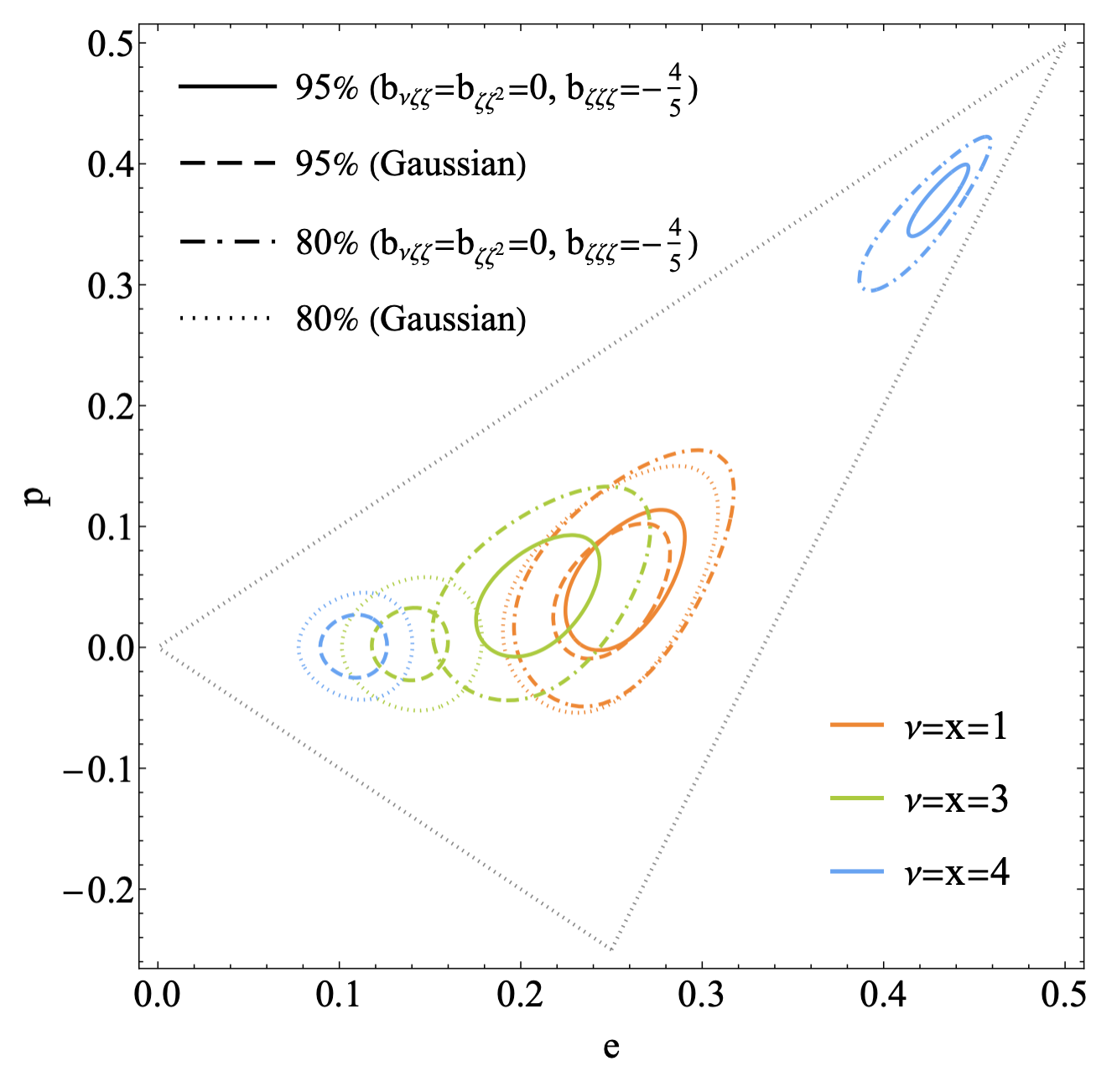}
    \caption{Comparison of the contour plots of $P_{\rm pk}(e,p|\nu,x)$ for fixed $\nu$ and $x$ between non-Gaussian and Gaussian random fields. The solid (dot-dashed) curves show $95\%\,(80\%)$ contour in the case with $b_{\nu\zeta\zeta}=b_{\zeta\zeta^{2}}=0\,,\,b_{\zeta\zeta\zeta} = -4/5$ and the dashed (dotted) curves show $95\%\,(80\%)$ contour in the case of Gaussian, respectively.}
    \label{fig:contourplot-comparison}
\end{figure}

As a numerical example, we show the contour plot of $P_{\rm pk}(e,p|\nu,x)$ for fixed $1\leq (\nu,x)\leq 4$ assuming $4\ll \min[{\nu}^c,{x}^c]$, with $b_{\nu\zeta\zeta}=b_{\zeta\zeta^{2}}=0\,,\,b_{\zeta\zeta\zeta} = -4/5$ in Fig.~\ref{fig:contourplot-ng}.  
We also show the comparison of contour plots of $P_{\rm pk}(e,p|\nu,x)$ between the case with $b_{\nu\zeta\zeta}=b_{\zeta\zeta^{2}}=0\,,\,b_{\zeta\zeta\zeta} = -4/5$ and the case of Gaussian random field discussed in Sec.~\ref{sec:Gaussian} in Fig.~\ref{fig:contourplot-comparison}. 
In this particular example, large non-Gaussianity makes the higher peaks less spherical.

One should note that the PDF is given by $\beta^{(1)}$ which has seven $b$ parameters (six independent $\tilde{\mathcal{I}}$ parameters) while $P_{\rm pk}(e,p|\nu,x)$ is given by only three of them. Therefore, it is non-trivial to connect the enhancement of the PDF for a tail in comparison with the Gaussian case as in Fig.~\ref{fig:schematic-pdf} and the trend of sphericity parameters, different from the case of local-type non-Gaussianities. We leave this investigation for future work.


\section{Summary and discussion}\label{sec:summary}

Peak theory plays crucial role to study the formation of large-scale structures and PBHs in the early universe. To first approximation, the statistics is expected to be Gaussian and most of the studies are restricted to the Gaussian statistics. However, deviation from the Gaussian case is inevitable in a real universe. We have provided the comprehensive analysis of the statistics of peaks of a random scalar field $F(\vec{r})$ taking the effect of non-Gaussianity into account. We then implemented our general framework to study the sphericity of peaks in the presence of non-Gaussianity. In the particular case of local-type non-Gaussianity, where $F$ is given by a functional of a Gaussian random field $F_{G}(\vec{r})$ such that $F=F[F_{G}]$, we have provided a general formalism which is applicable to any local-type non-Gaussianity no matter how large the deviation from Gaussian statistics is. For general non-Gaussianity, we provide a setup applicable to any power spectrum and bispectrum shape, 
neglecting higher-order correlators.

Through the above analysis, we have investigated how the shape around a peak is changed by the presence of non-Gaussianity. We present explicit expressions for the most probable values of the sphericity parameters, including the effect of non-Gaussianity on the shape. In the special case when the peaks of the non-Gaussian and Gaussian variables coincide, we have shown that rarer peaks tend to be higher and more likely to be spherical, similarly to the well-known result found in \cite{Bardeen:1985tr} for the Gaussian case. 
For general non-Gaussianity, the PDF is found from Edgeworth expansion, allowing for a similar analysis of peak theory of Gaussian random fields \cite{Bardeen:1985tr} to be applied to a random field with any shape of bispectrum. 
We found that the effects of perturbative non-Gaussianity on the sphericity parameters is negligible, as they are even smaller than the subleading Gaussian corrections. In other words, from statistical point of view, perturbative non-Gaussian effects have a negligible impact on peak sphericity.

Finally, we looked into the case of large non-Gaussianity. For local-type non-Gaussianity, in principle, large non-Gaussianity could make higher peaks tend to be less spherical, although we did not provide a concrete example of this behavior. Moreover, by focusing on the tail of the PDF, we have found that large non-Gaussianity can lead to much less spherical peak configurations in comparison with the Gaussian case, as illustrated by the concrete example shown in Figure 3. Therefore, when considering the effects of non-Gaussianity, the assumption of peak sphericity is not always sufficiently good. This is particularly important in the context of PBH formation, where the compaction function, which is based on the assumption of peak sphericity, plays a crucial role. We summarize the relationship among the tendency of height, rareness and sphericity of peaks in Table \ref{tab:summary}.

\begin{table}[]
    \centering
    \begin{tabular}{c||c|c}
         & Tendency of higher peaks & Tendency of rarer peaks \\ \hline\hline
         Perturbative non-Gaussianity & \multicolumn{2}{c}{\begin{tabular}{c}
                  One-to-one correspondence of the tendency \\
                  among three properties:\\
                  Higher $\leftrightarrow$ Rarer $\leftrightarrow$ More spherical
             \end{tabular}}\\ \hline
         \begin{tabular}{c}
                  Non-perturbative \\
                  Local-type non-Gaussianity
             \end{tabular}& No generic tendency & More spherical\\ \hline
         \begin{tabular}{c}
                  Non-perturbative \\
                  General bispectrum
             \end{tabular}& No generic tendency & No generic tendency
    \end{tabular}
    \caption{Summary of the relationship between the tendency of peak height, rareness, and sphericity in different regimes of non-Gaussianity}
    \label{tab:summary}
\end{table}

The origin of non-sphericity is an important aspect to investigate. For local-type non-Gaussianities, we demonstrated that rarer configurations tend to be more spherical, as the statistical properties such as the most probable values are determined by the Gaussian variable $F_{G}$ and its derivatives. However connecting the rareness and sphericity is non-trivial in general. In our analysis with general shape of bispectrum, we showed that a tail of the PDF depends on a larger set of parameters than the sphericity. As a result, linking the enhancement of the PDF for a tail in comparison with the Gaussian case and the trend of sphericity parameters is not straightforward. We leave this investigation for future work.

\section*{Acknowledgments}
MAG thanks organizers of the workshops ``IBS CTPU-CGA, Tokyo Tech, USTC 2024 summer school and workshop on cosmology, gravity, and particle physics" at Tateyama and ``COSMO'24” at Kyoto University, where this work was in its final stages. MAG also thanks Institute of Science Tokyo, Kavli Institute for the Physics and Mathematics of the Universe (IPMU), RIKEN Interdisciplinary Theoretical and Mathematical Sciences Program (iTHEMS), and Rikkyo University for hospitality and support during his visits. The work of MAG was supported by IBS under the project code, IBS-R018-D3. M.U. is supported by JSPS Grant-in-Aid for Research Fellows Grant No.24KJ1118 and by IBS under the project code, IBS-R018-D3. M.Y. is supported by IBS under the project code, IBS-R018-D3, and by JSPS Grant-in-Aid for Scientific Research Number JP23K20843. The research of CG is supported by the
grant PID2022-136224NB-C22, funded by MCIN\allowbreak/\allowbreak AEI\allowbreak/10.13039\allowbreak/501100011033\allowbreak/\allowbreak FEDER,
UE, and by the grant\allowbreak/ 2021-SGR00872.


\appendix

\section{Diagonalization of $\xi$}\label{app-xi-diag}
In this appendix, we demonstrate that for all cases of Gaussian, local non-Gaussianity, and small general non-Gaussianity, all quantities of interest in the corresponding PDF can be expressed in terms of $F$, $\eta_i = \nabla_i F$, and the three eigenvalues of $\xi_{ij} = \nabla_i \nabla_j F$, as the three other independent components of $\xi_{ij}$ become redundant after diagonalizing $\xi_{ij}$ and setting $\eta_i=0$.

\subsection{Gaussian}
The PDF is given by \eqref{PDF}. We first show that $Q$ depends only on $F$, $\eta_i$ and three eigenvalues of $\xi_{ij}$. After that, we look at the measure $\D{F}\D^3\eta\D^6\xi$ and show how we can integrate out the three independent components of $\xi$ other than the eigenvalues of $\xi$.

Using \eqref{eq:2 pt correlation} in \eqref{Q-def}, we find
\begin{align}\label{Q}
\begin{split}
2Q(F,\eta,\xi) &= \frac{1}{1-\gamma^{2}}
\left(
\frac{F^{2}}{\sigma_{0}^{2}} 
+ 2 \gamma \frac{F}{\sigma_0} \frac{\sum_{i}\xi_{ii}}{\sigma_2}
\right)
+
\frac{6-5\gamma^2}{1-\gamma^{2}} \frac{\sum_{i} \xi_{ii}^{2}}{\sigma_{2}^{2}} 
\\
&+ \frac{5\gamma^2-3}{1-\gamma^{2}}
\frac{\left( \xi_{11}\xi_{22} + \xi_{22}\xi_{33} + \xi_{33}\xi_{11}\right)}{\sigma_{2}^{2}}
+ \frac{15}{\sigma_{2}^{2}}\left(\xi_{12}^{2}+\xi_{23}^{2} +\xi_{13}^{2}\right)
+ \frac{3}{\sigma_{1}^{2}}\sum_{i}\eta_{i}^{2} \,,
\end{split}
\end{align}
where $\gamma= \sigma_{1}^{2}/\sigma_{0}\sigma_{2}$.  Let us look at the following quantities
\begin{align}\label{Traces}
\begin{split}
{\rm Tr}\left[\xi\right] 
&= \xi_{11} + \xi_{22} + \xi_{33} \,,
\\
{\rm Tr}\left[\xi^2\right]
&= \xi_{11}^2 + \xi_{22}^2 + \xi_{33}^2 
+ 2 \left( \xi_{12}^2 + \xi_{13}^2 + \xi_{23}^2 \right) \,,
\\
\frac{1}{2}\left({\rm Tr}\left[\xi\right]^2 -{\rm Tr}\left[\xi^2\right]\right) 
&= \xi_{11} \xi_{22} + \xi_{11} \xi_{33} + \xi_{22} \xi_{33}
- \left( \xi_{12}^2 + \xi_{13}^2 + \xi_{23}^2 \right) \,.
\end{split}
\end{align}
Using the above expression in \eqref{Q}, we find
\begin{align}\label{Q-Trace}
\begin{split}
2Q(F,\eta,\xi) =& 
\frac{1}{1-\gamma^{2}}
\left(
\frac{F^{2}}{\sigma_{0}^{2}} 
+ 2 \gamma \frac{F}{\sigma_0} \frac{{\rm Tr}\left[\xi\right]}{\sigma_2}
\right)
+ \frac{15}{2} \frac{{\rm Tr}\left[\xi^2\right]}{\sigma_2^2}
+\frac{5\gamma^2-3 }{2(1-\gamma^{2})} \frac{{\rm Tr}\left[\xi\right]^2}{\sigma_2^2}
+ \frac{3}{\sigma_{1}^{2}}\sum_{i}\eta_{i}^{2}  \,.
\end{split}
\end{align}
As it is clear from the above expression, $Q$ is completely determined with ${\rm Tr}\left[\xi\right]$ and ${\rm Tr}\left[\xi^2\right]$. This simple observation is quite important as we will see below.

Since $\xi$ is a symmetric matrix with real components, we can always diagonalize it
\begin{align}\label{trans}
&\xi =- R^{T}.\Lambda.R \,,
&& R.R^T = R^T.R = 1 \,,
\end{align}
such that 
\begin{align}
\Lambda = - R.\xi.R^{T} \doteq {\rm diag}(\Lambda_{1},\Lambda_{2},\Lambda_{3}) \,,
\end{align}
where $\Lambda_i$ are the eigenvalues of $\xi$. The negative sign is considered to work with positive eigenvalues $\Lambda_i>0$ as $\xi$ is negative definite for peaks. In what follows, our aim is to find a concrete expression for the PDF \eqref{PDF} after diagonalization. In other words, we substitute $\xi_{ij}$ and $(\D\xi)_{ij}$ in terms of six new variables which diagonalize $\xi$.

Let us first show that $Q(F,\eta,\xi)$ only depends on $\Lambda_i$ after diagonalization such that $Q(F,\eta,\xi)\to{Q}(F,\eta,\Lambda)$. Note that this is not trivial since in general there will be six new coordinates after diagonalization while we are claiming that only three $\Lambda_i$ will show up in $Q$. Indeed, after rewriting \eqref{Q} in the form \eqref{Q-Trace}, this is quite clear. We only need to note that traces \eqref{Traces} are completely characterized by the eigenvalues
\begin{align}
\begin{split}
{\rm Tr}\left[\xi\right] 
= - {\rm Tr}\left[R^T.\Lambda.R\right] 
= - {\rm Tr}\left[\Lambda.R^T.R\right] = - {\rm Tr}\left[\Lambda\right] 
&= - \left(\Lambda_1+\Lambda_2+\Lambda_3\right) \,,
\\
{\rm Tr}\left[\xi^2\right] 
= {\rm Tr}\left[R^T.\Lambda.R.R^T.\Lambda.R\right] 
= {\rm Tr}\left[R^T.\Lambda^2.R\right] 
= {\rm Tr}\left[\Lambda^2\right] 
&= \Lambda_1^2+\Lambda_2^2+\Lambda_3^2 \,,
\\
\frac{1}{2}\left({\rm Tr}\left[\xi\right]^2 -{\rm Tr}\left[\xi^2\right]\right)
&= \Lambda_1 \Lambda_2 + \Lambda_1 \Lambda_3 + \Lambda_2 \Lambda_3 \,.
\end{split}
\label{Trace-xi-UpTo2}
\end{align}

Using \eqref{Trace-xi-UpTo2} in \eqref{Q-Trace}, we find 
\begin{align}\label{Q-Lambda}
2Q\left(F,\eta,\Lambda\right) =& \frac{1}{1-\gamma^{2}}
\left( \frac{F^2}{\sigma_0^2} - 2 \gamma \frac{F\sum_i\Lambda_i}{\sigma_0\sigma_2} \right)
+ \frac{5\gamma^2-3}{2(1-\gamma^2)} \bigg(\frac{\sum_i \Lambda_i}{\sigma_2}\bigg)^2 + \frac{15}{2} \frac{\sum_i\Lambda_i^2}{\sigma_2^2}
+ 3 \frac{\sum_{i}\eta_{i}^{2} }{\sigma_{1}^{2}} \,,
\end{align}
which shows that $Q$ only depends on $F$, $\eta_i$ and eigenvalues $\Lambda_i$.

The next step is to find the Jacobian of transformation \eqref{trans} to find how measure $\D^6\xi$ changes after the diagonalization. This have been done in details in appendix B of \cite{Bardeen:1985tr}. The final result is 
\begin{align}\label{measure-Lambda}
\D^6\xi = \frac{1}{3!} |(\Lambda_{1}-\Lambda_{2})(\Lambda_{2}-\Lambda_{3})(\Lambda_{3}-\Lambda_{1})|\D\Lambda_{1}\D\Lambda_{2}\D\Lambda_{3}\D^3\Omega_{S^3},
\end{align}
where $\D^3\Omega_{S^3}$ is the volume element of a 3-sphere with $SO(3)$ symmetry group.

Substituting \eqref{Q-Lambda} and \eqref{measure-Lambda} in the PDF \eqref{PDF}, we find the PDF in terms of the diagonalized variables. After diagonalization, $\Lambda_i$ can be chosen as three new variables and the other three variables only show up through the volume element of the 3-sphere $\D^3\Omega_{S^3}$. Then, we can simply integrate the volume element which gives rise to the finite volume of the unit 3-sphere $\int\D^3\Omega_{S^3}=2\pi^2$.\footnote{While it is convenient to set up the Euler angles as coordinates on the 3-sphere, we do not need to do so since we only need the volume which is a global invariant of 3-sphere.} In this regard, we get rid of three coordinates (e.g. Euler angles) and find \eqref{PDF-lambda} in terms of three $\Lambda_i$ instead of six $\xi_{ij}$. That is why working with the diagonalized variables is much more convenient than the original variables $\xi_{ij}$.

\subsection{Local non-Gaussianity}\label{app-local-NG}

For the case of local non-Gaussianity, the PDF is given by \eqref{PDF-NG}. For the Gaussian case, we have shown that the dependency of $Q_G$ on $\xi_G$ can be completely expressed in terms of the three eigenvalues of $\xi_G$. This is proved through the diagonalization of $\xi_G$. Here we look for the implications of this result for the local non-Gaussian case. Taking trace of the last equation in \eqref{eq:hG of exact local nonG} we find
\begin{align}\label{Trace}
\sum_{i} {\Lambda_{G}}_i 
= J_1 \sum_{i} {\Lambda}_i - J_2 \sum_i \eta_i^2 \,,
\end{align}
where ${\Lambda_{G}}_i$ and $\Lambda_i$ are eigenvalues of $\xi_G$ and $\xi$, respectively. Taking trace of the square of $\xi_G$, we find
\begin{align}\label{Trace-2}
\sum_{i} {\Lambda_{G}}_i^2
= J_1^2 \sum_{i} {\Lambda}_i^2 + J_2^2 \Big(\sum_i \eta_i^2\Big)^2 - 2 J_1 J_2  \sum_{i} {\Lambda}_i \big(R_{ik}\eta_k\big)^2 \,,
\end{align}
where we have used
\begin{align}
- {\rm Tr}\left[\xi.\eta.\eta^T\right] 
= {\rm Tr}\left[R^T.\Lambda.R.\eta.\eta^T\right] 
= {\rm Tr}\left[\Lambda.R.\eta.\eta^T.R^T\right] 
= {\rm Tr}\left[(R.\eta)^T.\Lambda.(R.\eta)\right] \,.
\end{align}

With a similar and straightforward calculation, we find
\begin{align}\label{Trace-3}
\sum_{i} {\Lambda_{G}}_i^3
= J_1^3 \sum_{i} {\Lambda}_i^3 
- 3 J_1^2 J_2 \sum_{i} {\Lambda}_i^2 (R_{ik}\eta_k)^2
+ 3 J_1 J_2^2 \sum_{i} {\Lambda}_i \big(R_{ik}\eta_k\big)^2 \sum_j \eta_j^2
- J_2^3 \Big(\sum_i \eta_i^2\Big)^3
 \,.
\end{align}
From Eqs.~\eqref{Trace}, \eqref{Trace-2}, \eqref{Trace-3}, in principle, we can find ${\Lambda_{G}}_i$ in terms of $\Lambda_i$ and $\eta_i$ and after substituting ${{h}_G}_A$ in terms of $h_A$, $Q$ will be expressed in terms of not only the eigenvalues of $\xi$ and $\eta.\eta^T$ but also the three other independent components of $\xi$. Using Euler angles for the latter, this means, $Q$ depends on the Euler angles when $\eta\neq0$. However, $Q$ can be completely expressed in terms of eigenvalues of $\xi$ (e.g. independent of the Euler angles) for $\eta=0$ with which we are interested to study peaks.

\subsection{General non-Gaussianity}\label{app:genral-NG}
In this case, the PDF is given by \eqref{PDF-Edgeworth-2} with $K(F,\eta,\xi)$ is defined in \eqref{K-final}. Using  \eqref{kappas}, \eqref{kappa-xxx}, and \eqref{Q-Trace} in the definition of $\beta^{(1,2)}$ in Eq. \eqref{def-beta}, we find
\begin{align}
\begin{split}
\beta^{(1)}(\nu,\alpha,\varsigma) &= 
b_{\nu^{3}} \nu^3
+ b_{\nu^{2}\varsigma} \nu^2 {\rm Tr}(\varsigma) + b_{\nu\varsigma^2} \nu {\rm Tr}(\varsigma)^2
+ b_{\nu\varsigma\varsigma} \nu {\rm Tr}(\varsigma^2)
\\
&
+ b_{\varsigma\varsigma\varsigma} {\rm Tr}(\varsigma^3)
+ b_{\varsigma\varsigma^2} {\rm Tr}(\varsigma) {\rm Tr}(\varsigma^2)
+ b_{\varsigma^3} {\rm Tr}(\varsigma)^3
\\
&+ b_{\nu\alpha^2} \nu\alpha^2 + b_{\alpha^2\varsigma} \alpha^2 {\rm Tr}(\varsigma)
+ b_{\alpha\alpha\varsigma} {\rm Tr}\left(\alpha\alpha\varsigma\right)
\,,
\end{split}
\label{beta1-Trace}
\end{align}
and 
\begin{align}\label{beta2-Trace}
\beta^{(2)}(\nu,\varsigma) &= 
b_\nu \nu + b_\varsigma {\rm Tr}(\varsigma) \,,
\end{align}
where the normalized variables $(\nu,\alpha,\varsigma)$ are defined in \eqref{h-Normalized} and $\alpha=\sqrt{\delta_{ij}\alpha_i \alpha_j}$, ${\rm Tr}(\alpha\alpha \varsigma )= \alpha_i \alpha_j \varsigma_{ij}$. The explicit form of the coefficients are given by
\begin{align}
\begin{split}
b_{\nu^3} &= \frac{1}{6\left(1-\gamma^2\right)^3}
\Big[\tilde{\mathcal{I}}_0 + 3
\gamma ^2 \left(-6
	\tilde{\mathcal{I}}_1+15
	\tilde{\mathcal{I}}_2+9 \gamma  \tilde{\mathcal{I}}_3+45 \gamma 
	\tilde{\mathcal{I}}_4+35 \gamma  \tilde{\mathcal{I}}_5\right)
	\Big] \,,
\\
b_{\nu^2\varsigma} &=
\frac{\gamma }{2\left(1- \gamma ^2\right)^3}
\Big[\tilde{\mathcal{I}}_0
-6 \left(2 \gamma ^2+1\right)
	\tilde{\mathcal{I}}_1+15 \left(\gamma ^2+2\right)
	\tilde{\mathcal{I}}_2+9 \gamma  \left(\gamma ^2+2\right)
	\tilde{\mathcal{I}}_3+15 \gamma  \left(9 \tilde{\mathcal{I}}_4+7
	\tilde{\mathcal{I}}_5\right)
\Big] \,,
\\
b_{\nu\varsigma^2} &=
\frac{1}{4\left(1-\gamma^2\right)^3} \Big[2\gamma^{2}\tilde{\mathcal{I}}_{0} - 12\gamma^{2}(2+\gamma^{2})\tilde{\mathcal{I}}_{1}-15(3-14\gamma^{2}+5\gamma^{4})\tilde{\mathcal{I}}_{2}
\\
&\hspace{1cm}+18\gamma(1+2\gamma^{2})\tilde{\mathcal{I}}_{3} + 45\gamma(1+10\gamma^{2}-5\gamma^{4})\tilde{\mathcal{I}}_{4} - 105\gamma(3-10\gamma^{2} + 5\gamma^{4})\tilde{\mathcal{I}}_{5}
\Big] \,,
\\
b_{\nu\varsigma\varsigma} &= \frac{225}{4
\left(1-\gamma ^2\right)} \left(\tilde{\mathcal{I}}_2+3 \gamma  \tilde{\mathcal{I}}_4+7 \gamma 
\tilde{\mathcal{I}}_5\right) \,,
\\
b_{\varsigma\varsigma\varsigma} &= \frac{1125}{2} \tilde{\mathcal{I}}_5 \,,
\hspace{3cm}
b_{\varsigma\varsigma^2} = \frac{225}{4 \left(1-\gamma^2\right)}
\left[\left(10 \gamma ^2-3\right)
\tilde{\mathcal{I}}_5+\gamma  \tilde{\mathcal{I}}_2+3
\tilde{\mathcal{I}}_4\right] \,,
\\
b_{\varsigma^3} &=
\frac{1}{12\left(1-\gamma ^2\right)^3}
\Bigg[
2 \gamma ^3 \tilde{\mathcal{I}}_0 - 9 \gamma  \left(4 \gamma
	^2 \tilde{\mathcal{I}}_1+5 \left(5 \left(\gamma ^2-2\right)
	\gamma ^2+3\right) \tilde{\mathcal{I}}_2-6 \gamma 
	\tilde{\mathcal{I}}_3\right)
\\
&\hspace{2cm}
-135 \left(5 \left(\gamma ^2-2\right)
	\gamma ^2+3\right) \tilde{\mathcal{I}}_4-15 \left(100 \gamma
	^6-195 \gamma ^4+90 \gamma ^2-9\right) \tilde{\mathcal{I}}_5
\Bigg] \,,
\end{split}
\label{coeff-c}
\end{align}
\begin{align}
\begin{split}
b_\nu &= - \frac{1}{2\left(1-\gamma^2\right)^2}
\Big[
\tilde{\mathcal{I}}_0 + 3 \left(3-9 \gamma ^2\right) \tilde{\mathcal{I}}_1-45
\left(\gamma ^2-2\right) \tilde{\mathcal{I}}_2
\\
&\hspace{2cm}
+3 \gamma 
\left(\left(10 \gamma ^2-1\right) \tilde{\mathcal{I}}_3+15
\left(8-5 \gamma ^2\right) \tilde{\mathcal{I}}_4+35 \left(6-5
\gamma ^2\right)
\tilde{\mathcal{I}}_5\right)
\Big] \,,
\\
b_\varsigma 
&= - \frac{1}{2 \left(1-\gamma^2\right)^2}
\Big[
\gamma\tilde{\mathcal{I}}_{0} + 3\gamma(1-7\gamma^{2})\tilde{\mathcal{I}}_{1} + 3(-4 + 13\gamma^{2})\tilde{\mathcal{I}}_{3} 
\\
&\hspace{2cm}
 + 105(6-5\gamma^{2})\tilde{\mathcal{I}}_{5}+ 15(8-5\gamma^{2})\left(\gamma\tilde{\mathcal{I}}_{2} + 3\tilde{\mathcal{I}}_{4}\right)
\Big] \,,
\end{split}
\label{coeff-c-beta-2}
\end{align}
and
\begin{align}
\begin{split}
b_{\nu\alpha^2} &= \frac{3(3 \tilde{\mathcal{I}}_1-4 \gamma  \tilde{\mathcal{I}}_3)}{2(1-\gamma^2)}
\,,
\qquad 
b_{\alpha^2\varsigma} = \frac{3 [\left(5 \gamma ^2-9\right) \tilde{\mathcal{I}}_3+3\gamma  \tilde{\mathcal{I}}_1]}{2(1-\gamma^2)}
\,,
\qquad
b_{\alpha\alpha\varsigma} = \frac{45}{2} \tilde{\mathcal{I}}_3
\,,
\end{split}
\label{coeff-c-alpha}
\end{align}
where we have defined normalized dimensionless quantities
\begin{align}
\tilde{\mathcal{I}}_0 &\equiv \frac{{\cal I}_0}{\sigma_0^3} \,,
\qquad
\tilde{\mathcal{I}}_1
\equiv \frac{{\cal I}_1}{\sigma_0\sigma_1^2} \,,
\qquad
\tilde{\mathcal{I}}_2
\equiv \frac{{\cal I}_2}{\sigma_0\sigma_2^2} \,,
\qquad
\tilde{\mathcal{I}}_3
\equiv \frac{{\cal I}_3}{\sigma_1^2\sigma_2} \,,
\qquad
\tilde{\mathcal{I}}_4
\equiv \frac{{\cal I}_4}{\sigma_2^3} \,,
\qquad
\tilde{\mathcal{I}}_5
\equiv \frac{{\cal I}_5}{\sigma_2^3} \,.
\end{align}

In the results \eqref{beta1-Trace} and \eqref{beta2-Trace}, trace of powers of $\varsigma$ and ${\rm Tr}\left(\alpha\alpha\varsigma\right)$ showed up which are given by
\begin{align}\label{Tr-varsigma-n}
&{\rm Tr} \left[\varsigma^n\right]=(-1)^n\sum_i\lambda_i^n \,,
\qquad
{\rm Tr}\left(\alpha\alpha\varsigma\right) = \sum_i \lambda_i^2 \big(R_{ik}\alpha_k\big)^2 \,.
\end{align} 
Therefore, while \eqref{beta2-Trace} only depends on $\nu$ and three eigenvalues of $\varsigma$, \eqref{beta1-Trace} depends also on $\eta$ and the Euler angles (which characterize the three independent components of $\varsigma$ other than eigenvalues). Using this result together with \eqref{measure-Lambda}, we find that the non-Gaussian PDF \eqref{PDF-NG-lambda} depends on $\nu$, $\alpha_i$, the three eigenvalues of $\varsigma_{ij}$ and also Euler angles.

\section{Some useful functions}\label{app-functions}
In this appendix, we present the explicit forms of the functions $f(x)$ and $g_\beta(\nu,x)$ that are defined in Eqs. \eqref{def-f} and \eqref{def-g}, respectively.

Substituting \eqref{eq:Q(F,x) in Gaussian}, \eqref{eq:definition of J(e,p)}, and \eqref{eq:definition of chi(e,p)} in \eqref{def-f}, we find \cite{Bardeen:1985tr}
\begin{align}
\begin{split}
f({ x}) &= 
\frac{(5 { x}^2 - 16)}{3^2 5^4} e^{-\frac{5 { x}^2}{2}} 
+ \frac{\left(155
	{ x}^2+32\right)}{2\times 3^2 5^4} e^{-\frac{5 { x}^2}{8}} 
+ \frac{\sqrt{10 \pi } 
	\left({ x}^2-3\right)}{2\times 3^2 5^3} { x} \bigg(\text{erf}\bigg(\sqrt{\frac{5}{8}} { x}\bigg) + \text{erf}\bigg(\sqrt{\frac{5}{2}} { x}\bigg)\bigg) 
\,,
\end{split}
\label{function-f}
\end{align}
which have the following asymptotic behavior
\begin{align}
f({ x}) =
\begin{cases}
\frac{3^3}{5.7.2^{11}} { x}^8 \left(1-\frac{5 { x}^2}{8}\right) & { x} \to 0 \,,
\\ 
\frac{1}{5^23^2} \sqrt{\frac{2 \pi }{5}} \left({ x}^3-3{ x}\right) & { x} \to \infty \,.
\end{cases}
\label{function-f-limits}
\end{align}

Substituting \eqref{eq:Q(F,x) in Gaussian}, \eqref{def-beta-ep},  \eqref{eq:definition of J(e,p)}, and \eqref{eq:definition of chi(e,p)} in \eqref{def-g}, we find
\begin{align}
    \begin{split}
        g_{\beta}(\nu,x) &= \frac{2}{3^3 5^5}
        \big[
        \left(25
        { x}^2-128\right) \left( b_{\nu \varsigma\varsigma} \nu - b_{\varsigma \varsigma^2} x \right) + b_{\varsigma\varsigma\varsigma}
        \left(134-25 {x}^2\right) x
        \big] e^{-\frac{5 { x}^2}{2}} 
        \\
        &+ \frac{1}{2^3 3^3 5^5 }
        \big[
        2 \left(675 { x}^4+4180 { x}^2+1024\right) \left( b_{\nu \varsigma\varsigma}\nu -
        b_{\varsigma\varsigma^2} x \right) - b_{\varsigma\varsigma\varsigma} \left(1125 { x}^4+8180 { x}^2+464\right) x
        \big] e^{-\frac{5 { x}^2}{8}}
        \\
        &+ \frac{\sqrt{10 \pi }}{3^3 5^5} 
        \big[5(5x^2-21)x\left( b_{\nu \varsigma\varsigma}\nu -
        b_{\varsigma\varsigma^2} x \right) - b_{\varsigma\varsigma\varsigma} \left(25 { x}^4-105 { x}^2+14\right) 
        \big] \bigg(\text{erf}\bigg(
        \sqrt{\frac{5}{8}} { x}\bigg) + \text{erf}\bigg(\sqrt{\frac{5}{2}}
        { x} \bigg)\bigg) \,,
    \end{split}
    \label{function-g}
\end{align}
which has the asymptotic behavior
\begin{align}\label{asymptotic-g}
g_{\beta}(\nu,x) =
\begin{cases}
\frac{3^2}{5 .7.2^{12}} 
{ x}^{10}
\left[ b_{\nu \varsigma\varsigma} \nu
- \left( b_{\varsigma\varsigma^2}+\frac{73}{66} b_{\varsigma\varsigma\varsigma} \right) x \right] & { x} \to 0 \,,
\\ 
\frac{2}{3^3.5^2} \sqrt{\frac{2 \pi }{5}}
 x^3 \left[
b_{\nu \varsigma\varsigma} \nu
- \left( b_{\varsigma\varsigma^2} + b_{\varsigma\varsigma\varsigma}\right) x
\right] & { x} \to \infty \,.
\end{cases}
\end{align}

\section{Computation of $K(F,\eta,\xi)$ for $c^{(3)}_{ABC}\neq0$ \& $c^{(n\geq 4)}_{A_1\cdots{A}_n}=0$}\label{app-edgeworth}

Assuming that the 3-th cumulant is dominant and ignoring the 4-th and higher cumulants, \eqref{K-def-app-gen} simplifies to
\begin{align}\label{K-def-app}
	K(F,\eta,\xi) = 
	e^{Q_G} \exp\left(-\frac{1}{3!}\sum_{A,B,C} \beta_{ABC} \partial_{h_A} \partial_{h_B} \partial_{h_C} \right)
	e^{-Q_G} \,,
\end{align}
where 
\begin{align}
	\beta_{ABC} \equiv c^{(3)}_{ABC} = \langle h_{A}h_{B}h_{C} \rangle \,.
	\label{def-betas-app}
\end{align}
To operate the derivatives in \eqref{K-def-app}, we expand the exponential
\begin{align}
	K(F,\eta,\xi) = e^{Q_G} 
	\bigg[
	1 - \frac{1}{3!} \beta_{ABC} \partial_{A} \partial_{B} \partial_{C} 
	+ \frac{1}{(3!)^2} \beta_{ABC} \beta_{DEF} \partial_{A} \partial_{B} \partial_{C} \partial_{E} \partial_{D} \partial_{F} 
	+ \cdots 
	\bigg]
	e^{-Q_G} \,,
	\label{def-K}
\end{align} 
where $\cdots$ denotes terms that are higher order in $\beta_{ABC}$.
We note that even if we have restricted ourselves to the case that the 3-th cumulant dominates over all higher order cumulants, still we need to look at all powers of $\beta_{ABC}$ in \eqref{def-K}. Therefore, in the following, we find closed forms for $K(F,\eta,\xi)$ for two cases of small non-Gaussianity with ${\cal O}(\beta_{ABC})\ll1$ and focusing on the tail $h_A \gg \sigma_A$ where $\sigma_{F}=\sigma_0, \sigma_{\eta_i} = \sigma_1, \sigma_{\xi_{ij}} = \sigma_2$ such that $\nu, \alpha, \varsigma$ coincide with their definitions in \eqref{h-Normalized}.

\subsection{Small non-Gaussianity up to the cubic order}\label{subsec:derivation of NG PDF}

Assuming that ${\cal O}(\beta_{ABC})\ll1$, we can safely ignore ${\cal O}(\beta_{ABC}^n)$ with $n\geq2$ in \eqref{def-K}. Then, taking into account the fact that
\begin{align}
	\partial_{A} Q_G = M^{-1}_{AB} h_B \,,
	\qquad
	\partial_{A} \partial_{B} Q_G = M^{-1}_{AB} \,,
	\qquad 
	\partial_{A} \partial_{B} \partial_{C} Q_G = 0 \,,
	\label{D-Q-b}
\end{align}
which is clear from \eqref{def-QG}, \eqref{def-K} simplifies to 
\begin{align}\label{K-final-app}
	K(F,\eta,\xi) &\simeq 1 + \beta^{(1)}(F,\eta,\xi) + \beta^{(2)}(F,\xi) \,,
\end{align}
where we have defined
\begin{align}\label{def-beta-app}
	\begin{split}
		\beta^{(1)}(F,\eta,\xi) &\equiv  \frac{1}{3!} \beta_{ABC} \partial_{A}Q_G \partial_{B}Q_G \partial_{C}Q_G = 
		\frac{1}{3!} \beta_{ABC} M^{-1}_{AD} M^{-1}_{BE} M^{-1}_{CF} h_D h_E h_F \,,
		\\
		\beta^{(2)}(F,\xi) &\equiv - \frac{1}{2} \beta_{ABC} \partial_{A}Q_G \partial_{B} \partial_{C}Q_G =
		- \frac{1}{2} \beta_{ABC} M^{-1}_{AD} M^{-1}_{BC} h_D
		\,,
	\end{split}
\end{align}
in which we have used \eqref{D-Q-b} in the last steps.

\subsection{Tail behavior}
We now relax the assumption  ${\cal O}(\beta_{ABC})\ll1$ and consider ${\cal O}(\beta_{ABC})={\cal O}(1)$. In this case, we cannot ignore higher orders of $\beta_{ABC}$ in \eqref{def-K}. 

Using \eqref{D-Q-b}, \eqref{def-K} simplifies to 
\begin{align}
	\begin{split}
		K(F,\eta,\xi) &= 1 
		+ \frac{1}{3!} \beta_{ABC} \partial_{A}Q_G \partial_{B}Q_G \partial_{C}Q_G
		- \frac{1}{2} \beta_{ABC} \partial_{A}Q_G \partial_{B} \partial_{C}Q_G
		\\
		& + \frac{1}{2!} \Big( \frac{1}{3!} \beta_{ABC} \partial_{A}Q_G \partial_{B}Q_G \partial_{C}Q_G \Big)^2
		\\
		& - \frac{1}{24} \left(  2\beta_{ABC} \beta_{DEF} + 3 \beta_{ABE} \beta_{CDF} \right) \partial_{A}Q_G \partial_{B}Q_G \partial_{C}Q_G \partial_{D}Q_G
		\partial_{E}\partial_{F}Q_G 
		\\
		& + \frac{1}{8} \left( \beta_{ACD} \beta_{BEF} + 2 \beta_{ACE} \beta_{BDF} + 2 \beta_{ABC} \beta_{DEF} \right) \partial_{A}Q_G \partial_{B}Q_G \partial_{C} \partial_{D}Q_G
		\partial_{E}\partial_{F}Q_G 
		\\
		& - \frac{1}{24} \left( 2 \beta_{ACE} \beta_{BDF} + 3 \beta_{ABC} \beta_{DEF} \right) \partial_{A} \partial_{B}Q_G \partial_{C}\partial_{D}Q_G
		\partial_{E}\partial_{F}Q_G + \cdots \,.
	\end{split}
	\label{K-series}
\end{align}
From \eqref{D-Q-b} we see that $\partial_{A} Q_G$ is linear in $h_A$ while $\partial_{A} \partial_{B} Q_G$ is independent of $h_A$. Consequently, the first is larger/smaller than the latter for $(h_A\to\infty)/(h_A\to0)$. Thus, in the series \eqref{K-series} those terms which include the highest power of $\partial_{A} Q_G$ are dominant for $h\to\infty$ while those which include the highest power of $\partial_{A} \partial_{B} Q_G$ are dominant for $h\to0$. In order to better see this claim, it is useful to work with the normalized quantities
\begin{align}\label{def-h-tilde}
	{\tilde h}_A \equiv \frac{h_A}{\sigma_{h_A}} 
	= \{ \nu, \alpha, \varsigma \} \,,
	\qquad 
	{\tilde \partial}_A \equiv \partial_{{\tilde h}_A} = \sigma_{A}\, \partial_{A} \,.
\end{align}
Working with \eqref{def-h-tilde} it is reasonable to define the following quantities
\begin{align}
	\tilde{\partial}_{A} \tilde{\partial}_{B} Q_G &\equiv \tilde{M}^{-1}_{AB} = \sigma_{h_A} \sigma_{h_B} {M}^{-1}_{AB}
	\,,
	\qquad
	{\tilde \beta}_{ABC} \equiv \frac{ \beta_{ABC}}{\sigma_{h_A}\sigma_{h_B}\sigma_{h_C}} \,.
\end{align}
We thus have $Q_G=\tilde{M}^{-1}_{AB}\tilde{h}_A\tilde{h}_B$ and from \eqref{Q-Trace} we see that $\tilde{M}^{-1}_{AB}={\cal O}(1)$. To keep the expansion \eqref{K-series} in the perturbative regime, we need to assume that all terms in the expansion are smaller than unity. This leads to the following result $\tilde{\beta}_{\max} = \min\left[ \tilde{h}^{-3}, \tilde{h}^{-2}, \tilde{h}^{-1}, 1 \right]$ where $\tilde{\beta}$ and $\tilde{h}$ schematically show the order of $\tilde{\beta}_{ABC}$ and $\tilde{h}_A$. If we compare the last term in the first line of \eqref{K-series}, which is first order ${\cal O}({\tilde\beta})$, with the second order ${\cal O}({\tilde\beta}^2)$ terms in the last line of \eqref{K-series}, we surprisingly find that for $\tilde{\beta} > \tilde{h}$, the second order term ${\cal O}\big(\tilde{\beta}^2\big)$ dominates the first order term ${\cal O}\big(\tilde{\beta}\big)$. This means that, to ensure that always all first order terms ${\cal O}\big(\tilde{\beta}\big)$ dominate over the second order terms ${\cal O}\big(\tilde{\beta}^2\big)$, we should work in the regime $\tilde{\beta} \ll \tilde{h}$. Therefore, the result \eqref{K-final-app} is only valid when we take into account that $\tilde{\beta}\ll\tilde{\beta}_{\max}$ defined as
\begin{align}\label{kappa-max}
	\tilde{\beta}_{\max} = \min\left[ \tilde{h}^{-3}, \tilde{h}^{-2}, \tilde{h}^{-1}, 1, \tilde{h} \right] \,.
\end{align}

Let us now look at the tail of the PDF $h\to\infty$. More precisely, we deal with the limit $h_A \gg \sigma_{A}$ which is equivalent to $\tilde{h}_A\gg1$. In order to do so, we rewrite \eqref{K-series} in the following form
\begin{align}
	\begin{split}
		K(F,\eta,\xi) &= \left[ 1 
		+ \frac{1}{3!} \beta_{ABC} \partial_{A}Q_G \partial_{B}Q_G \partial_{C}Q_G
		+ \frac{1}{2!} \Big( \frac{1}{3!} \beta_{ABC} \partial_{A}Q_G \partial_{B}Q_G \partial_{C}Q_G \Big)^2
		+ \cdots \right]
		\\
		& - \frac{1}{2} \beta_{ABC} \partial_{A}Q_G \partial_{B} \partial_{C}Q_G
		\\
		& - \frac{1}{24} \left(  2\beta_{ABC} \beta_{DEF} + 3 \beta_{ABE} \beta_{CDF} \right) \partial_{A}Q_G \partial_{B}Q_G \partial_{C}Q_G \partial_{D}Q_G
		\partial_{E}\partial_{F}Q_G 
		\\
		& + \frac{1}{8} \left( \beta_{ACD} \beta_{BEF} + 2 \beta_{ACE} \beta_{BDF} + 2 \beta_{ABC} \beta_{DEF} \right) \partial_{A}Q_G \partial_{B}Q_G \partial_{C} \partial_{D}Q_G
		\partial_{E}\partial_{F}Q_G 
		\\
		& - \frac{1}{24} \left( 2 \beta_{ACE} \beta_{BDF} + 3 \beta_{ABC} \beta_{DEF} \right) \partial_{A} \partial_{B}Q_G \partial_{C}\partial_{D}Q_G
		\partial_{E}\partial_{F}Q_G 
		+ \cdots \,.
	\end{split}
	\label{K-series-2}
\end{align}
In the limit $\tilde{h}_A\gg1$, the terms in the first line dominate and we can ignore all other terms. We can then sum up the series with an exponential as
\begin{align}\label{K-final-exp}
	K(F,\eta,\xi) \simeq e^{\beta^{(1)}(F,\eta,\xi)}
	\qquad 
	h_A \gg \sigma_{A} \,,
\end{align}
where $\beta^{(1)}(F,\eta,\xi)$ is defined in Eq. \eqref{def-beta-app}. 

\section{Computation of $I_{ABC}$}\label{app-I-ABC}
The three-point correlations $\beta_{ABC}$ are fixed by six parameters $\mathcal{I}_{I}(I=0\cdots 5)$, which are given by integrating the bispectrum $B(k,k',\theta)$ and bispectrum-independent parameter $I_{I}$ as in Eq.~\eqref{functions-I}. 
In this appendix, we present the detail derivation of Eq.~\eqref{functions-I}.

Take the two 3-dimensional momenta $\vec{k}_{1}$ and $\vec{k}_{2}$ as
\begin{align}
    \begin{split}
        \vec{k}_{1}&=k_{1}R_{z}(\phi_{1})R_{y}(\theta_{1})\begin{pmatrix}0\\0\\1\end{pmatrix}\,,\quad
        \vec{k}_{2}=k_{2}R_{z}(\phi_{1})R_{y}(\theta_{1})\begin{pmatrix}\sin{\theta}\cos{\phi}\\\sin{\theta}\cos{\phi}\\\cos{\theta}\end{pmatrix}\,,
    \end{split}\label{eq:vector k}
\end{align}
where
\begin{align}
    \begin{split}
        R_{z}(\alpha)=\begin{pmatrix}
            \cos{\alpha}&-\sin{\alpha}&0\\\sin{\alpha}&\cos{\alpha}&0\\0&0&1
        \end{pmatrix}\,,\quad 
        R_{y}(\alpha)=\begin{pmatrix}
            \cos{\alpha}&0&\sin{\alpha}\\0&1&0\\-\sin{\alpha}&0&\cos{\alpha}
        \end{pmatrix}\,,
    \end{split}
\end{align}
are rotational matrices with an angle $\alpha$ around $z$- and $y$- axes, respectively. 
Due to the homogeneity of the Universe, the three 3-momenta in the three-point correlation $\langle F(\vec{k}_{1}),F(\vec{k}_{2}),F(\vec{k}_{3})\rangle$ satisfy $\sum \vec{k}_{i}=0$. Thus, $k_{3}=|\vec{k}_{3}|$ is a function of $k_{1}$ $k_{2}$, and $\theta$.
Then, the 3-point correlation is reduced to
\begin{align}
    \begin{split}
        \beta_{ABC}
        =&\frac{1}{8\pi^{4}}\int \D k_{1}\D k_{2}\D \cos{\theta} \,k_{1}^{2}k_{2}^{2}B(k_{1},k_{2},\theta)I_{ABC}(k_{1},k_{2},\theta)\\
        =&\int \frac{\D^{3}{k_{1}}}{(2\pi)^{3}}\int \frac{\D^{3}{k_{2}}}{(2\pi)^{3}} B(k_{1},k_{2},\theta)I_{ABC}(k_{1},k_{2},\theta)\,,
    \end{split}
\end{align}
where
\begin{align}
    I_{ABC}(k_{1},k_{2},\theta)=\frac{1}{2(2\pi)^{2}}\int \D\phi \D\phi_{1} \D\cos{\theta_{1}}\,f_{ABC}(k_{1},k_{2},\theta,\theta_{1},\phi,\phi_{1})\,,\label{eq:I_ABC}
\end{align}
is independent of the specific shape of the bispectrum. The function $f_{ABC}$ is given by,
\begin{align}
    \begin{split}
        f_{FFF}&=1\,,\quad 
        f_{F\eta_{i}\eta_{j}}=-\frac{1}{3}({k_{1}}_{i}{k_{2}}_{j}+{k_{2}}_{i}{k_{3}}_{j}+{k_{3}}_{i}{k_{1}}_{j})\,,\quad
        f_{F^{2}\xi_{ij}}=-\frac{1}{3}({k_{1}}_{i}{k_{1}}_{j}+{k_{2}}_{i}{k_{2}}_{j}+{k_{3}}_{i}{k_{3}}_{j})\,,\\
        f_{\eta_{i}\eta_{j}\xi_{kl}}&=\frac{1}{3}({k_{1}}_{i}{k_{2}}_{j}{k_{3}}_{k}{k_{3}}_{l}+{k_{2}}_{i}{k_{3}}_{j}{k_{1}}_{k}{k_{1}}_{l}+{k_{3}}_{i}{k_{1}}_{j}{k_{2}}_{k}{k_{2}}_{l})\,,\\
        f_{F\xi_{ij}\xi_{kl}}&=\frac{1}{3}({k_{1}}_{i}{k_{1}}_{j}{k_{2}}_{k}{k_{2}}_{l}+{k_{2}}_{i}{k_{2}}_{j}{k_{3}}_{k}{k_{3}}_{l}+{k_{3}}_{i}{k_{3}}_{j}{k_{1}}_{k}{k_{1}}_{l})\,,\\
        f_{\xi_{ij}\xi_{kl}\xi_{mn}}&=-\frac{1}{3}({k_{1}}_{i}{k_{1}}_{j}{k_{2}}_{k}{k_{2}}_{l}{k_{3}}_{m}{k_{3}}_{n}+{k_{2}}_{i}{k_{2}}_{j}{k_{3}}_{k}{k_{3}}_{l}{k_{1}}_{m}{k_{1}}_{n}+{k_{3}}_{i}{k_{3}}_{j}{k_{1}}_{k}{k_{1}}_{l}{k_{2}}_{m}{k_{2}}_{n})\,,
    \end{split}
\end{align}
where the indices $i,j,k,l,m,n$ run from $1$ to $3$ which label the components of the momenta $\vec{k}_{1}$ and $\vec{k}_{2}$ as in Eq.~\eqref{eq:vector k}. It is straightforward to find $I_{\rm ABC}$ from Eq.~\eqref{eq:I_ABC}. The results are
\begin{align}
    \begin{split}
        I_{F^{3}} &= 1\,,\quad 
        I_{F\eta_{i}\eta_{j}}=-\frac{1}{2}I_{F^{2}\xi_{ij}}=\frac{1}{9}(k_1^2 + k_2^2 + k_1 k_2 \cos{\theta})\delta_{ij}\,,\\
        I_{F\xi_{ij}^{2}}&=
        \begin{cases}
            \frac{1}{15} ((k_1^2 + k_2^2)^2 + k_1 k_2 (2 (k_1^2 + k_2^2) \cos{\theta}+ k_1 k_2 \cos{2\theta}))\,,\quad(i=j)\,\\
            \frac{1}{180}(4k_{1}^{4} + 3k_{1}^{2}k_{2}^{2}+ 4k_{2}^{4} + k_1 k_2 (8 (k_1^2 + k_2^2) \cos{\theta}+9k_1 k_2 \cos{2\theta}))\,,\quad (i\neq j)\,
        \end{cases}
        \\
        I_{F\xi_{ii}\xi_{jj}} &=\frac{1}{90} (2k_{1}^{4} + 9k_{1}^{2}k_{2}^{2}+ 2k_{2}^{4} + k_1 k_2 (4 (k_1^2 + k_2^2) \cos{\theta}-3k_1 k_2 \cos{2\theta}))\,,\quad (i\neq j)\\
        I_{\eta_{i}\eta_{i}\xi_{jj}}&=-\frac{1}{2}I_{\eta_{i}\eta_{j}\xi_{ij}}=-\frac{1}{9} k_{1}^2 k_2^2 \sin^2{\theta}\,,\quad (i\neq j)\,
    \end{split}
\end{align}
and
\begin{align}
    \begin{split}\label{eq:I_xi^3}
        I_{\xi_{ii}^{3}}
        =&-\frac{1}{35} k_1^2 k_2^2(9 k_1 k_2 \cos{\theta}+ (k_1^2 +k_2^2) (3 + 2 \cos{2\theta}) + k_1 k_2 \cos{3\theta})\,,\\
        I_{\xi_{ii}^{2}\xi_{jj}}
        =&\frac{1}{315} k_1^2 k_2^2 (-19 k_1 k_2 \cos{\theta}+ (k_1^2 +k_2^2) (-11 + 2 \cos{2\theta}) + k_1 k_2 \cos{3\theta})\,,\\
        I_{\xi_{ij}^{2}\xi_{ii}}
        =&-\frac{1}{630} k_1^2 k_2^2  (31 k_1 k_2 \cos{\theta}+2 (k_1^2 +k_2^2) (4 + 5 \cos{2\theta}) +5 k_1 k_2 \cos{3\theta})\,,\\
        I_{\xi_{11}\xi_{22}\xi_{33}}
        =&\frac{1}{210} k_1^2 k_2^2  (-5 k_1 k_2 \cos{\theta}+2 (k_1^2 +k_2^2) (-2 +  \cos{2\theta}) + k_1 k_2 \cos{3\theta})\,,\\
        I_{\xi_{ij}^{2}\xi_{kk}}
        =&-\frac{1}{1260}k_1^2 k_2^2  (23 k_1 k_2 \cos{\theta}+2 (k_1^2 +k_2^2) (5 +  \cos{2\theta}) + k_1 k_2 \cos{3\theta})\,,\\
        I_{\xi_{12}\xi_{23}\xi_{31}}
        =&-\frac{1}{840} k_1^2 k_2^2  (13 k_1 k_2 \cos{\theta}+2 (k_1^2 +k_2^2) (1 +  3\cos{2\theta}) + 3k_1 k_2 \cos{3\theta})\,,
    \end{split}
\end{align}
where $i\neq j, \,j\neq k$ and $i\neq k$ in Eq.~\eqref{eq:I_xi^3}.
Therefore, $I_{ABC}$ are parameterized by six parameters $I_{I}(I=0\cdots 5)$ as
\begin{align}
    \begin{split}
        I_{F^{3}}&=I_{0}\,,\quad I_{F\eta_{i}\eta_{j}}=I_{1}\delta_{ij}\,,\\
        I_{F\xi_{ij}\xi_{kl}} &= I_{2}(\delta_{ij}\delta_{kl} + \delta_{ik}\delta_{jl} + \delta_{il}\delta_{jk}) +I_{3}\delta_{ij}\delta_{kl}\,,\quad
        I_{\eta_{i}\eta_{j}\xi_{kl}}=\frac{1}{3}I_{3}(-2\delta_{ij}\delta_{kl} + \delta_{ik}\delta_{jl} + \delta_{il}\delta_{jk})\,,\\
        I_{\xi_{ij}\xi_{kl}\xi_{mn}} 
        &=
        I_4
        \big[
        3\delta_{ij} \delta_{kl} \delta_{mn} +\delta_{ij} (\delta_{kn} \delta_{lm} + \delta_{km} \delta_{ln})
        + (\delta_{in} \delta_{jm} + \delta_{im} \delta_{jn}) \delta_{kl} + (\delta_{il} \delta_{jk} + \delta_{ik} \delta_{jl}) \delta_{mn} \big]
        \\
        &+
        I_5 \big[
        \delta_{il} (\delta_{jn} \delta_{km} + \delta_{jm} \delta_{kn}) 
        + (\delta_{ik} \delta_{jn} + \delta_{ij} \delta_{kn} ) \delta_{lm} + \delta_{in} (\delta_{jm} \delta_{kl} + \delta_{jl} \delta_{km} + \delta_{jk} \delta_{lm}) 
        \\
        &+ (\delta_{ik} \delta_{jm} + \delta_{ij} \delta_{km} ) \delta_{ln} + \delta_{im} (\delta_{jn} \delta_{kl} + \delta_{jl} \delta_{kn} + \delta_{jk} \delta_{ln}) + (\delta_{il} \delta_{jk} + \delta_{ik} \delta_{jl} + \delta_{ij} \delta_{kl}) \delta_{mn} 
        \big] \,,
    \end{split}
\end{align}
such that
\begin{align}
    \begin{split}
        I_{0}&=I_{F^{3}}\,,\quad
        I_{1}=I_{F\eta_{i}^{2}}\,,\quad 
        I_{2}=I_{F\xi_{ij}^{2}}(i\neq j)\,,\quad 
        I_{3}=-\frac{3}{2}I_{\eta_{i}^{2}\xi_{ij}}\,\\
        I_{4}&= I_{\xi_{ii}\xi_{jk}^{2}}-I_{\xi_{12}\xi_{23}\xi_{31}}\,,\quad I_{5}=I_{\xi_{12}\xi_{23}\xi_{31}}\,.
    \end{split}
\end{align}
From the above results one can easily find Eq.~\eqref{functions-I}.

\section{Result with specific local-type non-Gaussianity}\label{sec:specific bispectrums}
In Sec.~\ref{sec:exact local nonG} and Sec.~\ref{sec:general nonG} we have developed two different formalisms to take into account the roles of non-Gaussianity in the statistics of a random field. In this appendix, as a consistency check, we apply both of these formalisms to the specific local-type non-Gaussianity and we confirm that the results of Sec.~\ref{sec:exact local nonG} and Sec.~\ref{sec:general nonG} coincide in the regime of small (perturbative) non-Gaussianity.

Assuming non-Gaussianity is small, it is convenient to consider the map $F=F[F_{G}]$, that is defined in \eqref{eq:definition of local type F}, as $F=F_{G} + f_{\rm NL}F_{G}^{2} +\cdots$, where $f_{\rm NL}$ is a free parameter that characterizes the amplitude of bispectra and so on. In this case, we can apply both formalisms developed in Sec.~\ref{sec:exact local nonG} and Sec.~\ref{sec:general nonG}. Thus, to be more concrete, let us consider \eqref{eq:fNL expansion of F-L} which deals with the following simple expression\footnote{We have ignored the conventional prefactors, i.e. $3/5$, in front of $f_{\rm NL}$ for the sake of simplicity.}
\begin{align}
F = F_{G} + f_{\rm NL}(F_{G}^{2}-\sigma_{G,0}^{2}) \,.
\label{eq:fNL expansion of F}
\end{align}
The constant $\sigma_{G,0}$ is introduced for the convenience. It plays the role of Gaussian variance in the regime $f_{\rm NL}\to0$ or more precisely $|f_{\rm NL}|\ll1/\sigma_{G,0}$.

Let us first apply the formalism we have developed in Sec.~\ref{sec:exact local nonG}. For the local map \eqref{eq:fNL expansion of F}, we have
\begin{align}\label{Jaobian-local-specific}
J_{1}(F) = \frac{1}{\sqrt{1+4 f_{\rm{NL}}\left(F+\sigma _{G,0}^2 f_{\rm{NL}}\right)}} \,.
\end{align}
Substituting it together with \eqref{eq:fNL expansion of F} in \eqref{PDF-NG}, we find the joint PDF $P(F,\eta,\xi)$. We can also find the peak distribution function \eqref{eq:pdf (peaks) of Gaussian (ep)-NG}. For the sphericity parameter, Eq. \eqref{eq:em of exact local nonG with fixed Fx} gives
\begin{align}\label{em-L-NG}
e_{m}^{-2} = 6+\frac{5X^{2}}{\sigma_{G,2}^{2}\left[1+4 f_{\rm{NL}}\left(F+\sigma _{G,0}^2 f_{\rm{NL}}\right)\right]} \,,
\end{align}
while $p_{m}$ does not change. The above result is exact in the sense that it is valid for any value of $f_{\rm NL}$.

Now, let us apply our general formalism that is presented in Sec.~\ref{sec:general nonG}. The starting point is to find the bispectrum \eqref{bispectrum}. For \eqref{eq:fNL expansion of F}, the bispectrum can be expressed in terms of power spectra as
\begin{align}
\begin{split}
B(k_{1},k_{2},\theta)=2f_{\rm NL}  \left[ P(k_{1})P(k_{2})+P(k_{2})P(k_{3}(k_{1},k_{2},\theta))+P(k_{3}(k_{1},k_{2},\theta))P(k_{1}) 
\right] \,,
\end{split}
\end{align}
where $k_{3}=|-\vec{k}_{1}-\vec{k}_{2}|$ and $\theta=\cos^{-1}(\vec{k}_{1}\cdot\vec{k}_{2}/k_{1}k_{2}$).
Substituting the above bispectrum in Eq.~\eqref{def-cal-I}, we find the explicit forms of $\cal{I}_{I}$ in terms of the two-point correlations $\sigma_{j}$ defined in \eqref{sigma-def}. Doing so, the normalized coefficients ${\tilde{\cal I}}_{I}$ in \eqref{def-I-N} are obtained as
\begin{align}
\begin{split}
{\tilde{\cal I}}_{0} = \frac{9}{2} {\tilde{\cal I}}_{1} = \frac{45}{2} {\tilde{\cal I}}_{2} = 6 f_{\rm NL}\sigma_{0} \,, \qquad
{\tilde{\cal I}}_{3} = - \frac{15}{2} {\tilde{\cal I}}_{4} = \frac{2}{3}f_{\rm NL} \frac{\sigma_{1}^{2}}{\sigma_2} \,,\qquad 
{\tilde{\cal I}}_{5}=0\,.
\end{split}
\label{I-Local-NG}
\end{align}
Substituting the above results in \eqref{coeff-c}, \eqref{coeff-c-beta-2}, and \eqref{coeff-c-alpha}, we find the explicit forms of the coefficients $b$ as
\begin{align}
\begin{split}
b_{\nu^{3}} &=\frac{f_{\rm NL}\sigma_{0}}{1-\gamma^{2}}\,, \qquad b_{\nu^{2}\varsigma} = \frac{3f_{\rm NL}\gamma\sigma_{0}}{1-\gamma^{2}}\,,\qquad b_{\nu\alpha\alpha} = \frac{2f_{\rm NL}(3-2\gamma^{2})\sigma_{0}}{1-\gamma^{2}}\,,\\
b_{\alpha^{2}\varsigma}&=\gamma b_{\nu\varsigma^{2}} = \frac{f_{\rm NL}(5\gamma^{2}-3)\gamma\sigma_{0}}{1-\gamma^{2}}\,,\qquad b_{\alpha\alpha\varsigma} = \gamma b_{\nu\varsigma\varsigma} = 15f_{\rm NL}\gamma\sigma_{0}\,,\\
b_{\varsigma^{3}}&=b_{\varsigma\varsigma^{2}} = b_{\varsigma\varsigma\varsigma} = 0 \,,
\end{split}
\end{align}
which are valid up to linear order in $f_{\rm NL}$. Having the explicit forms of the coefficients $b$, we can find the joint PDF Eq.~\eqref{PDF-Edgeworth-2} for the local-type non-Gaussianity \eqref{eq:fNL expansion of F} in the perturbative regime. From Eq.~\eqref{eq:em,pm general}, $E_{m}=\{e_{m},p_{m}\}$ for fixed $\nu$ and large fixed $x$ are given as
\begin{align}\label{em-G-NG}
\begin{split}
e_{m}^{-2}&=6+5x^{2}(1-4f_{\rm NL}\nu\sigma_{0})\,,\qquad 
p_{m}=6e^{4}_{m}\,.
\end{split}
\end{align}
The asymptotic behavior of sphericity parameters for $\nu\gg1$ can be easily found by substituting \eqref{I-Local-NG} in Eq.~\eqref{eq:asymptotic e,p for general nonG}, namely,
\begin{align}
\begin{split}
e_{m}^{-2}\simeq 5\gamma^{2}\nu^{2}+ 6 + 30(1-\gamma^{2}) - 10f_{\rm NL}\sigma_{0}\gamma^{2}\nu^{3}\,.
\end{split}
\end{align}

Expanding \eqref{em-L-NG} 
for $f_{\rm NL}\ll1$ and using the fact that $\sigma_{G,i}\approxeq \sigma_{i}$ up to the first order in $f_{\rm NL}$, we find that \eqref{em-L-NG} 
coincide to \eqref{em-G-NG} up to the first order in $f_{\rm NL}$. 
Note that the above results are for positive $x$, but as discussed in the previous section, it is easy to obtain consistent results for negative $x$.
Thus, up to the first order in $f_{\rm NL}$, all results of this section can be equivalently found from either Sec.~\ref{sec:exact local nonG} or Sec.~\ref{sec:general nonG}. This shows that our different setups in Sec.~\ref{sec:exact local nonG} and Sec.~\ref{sec:general nonG} are consistent with each other.

\bibliography{reference} 
\bibliographystyle{JHEP}

\end{document}